\documentclass[useAMS,usenatbib,usegraphicx]{mn2e}

\title[Lyman `bump' galaxies]{Lyman `bump' galaxies - II. A possible
signature of massive extremely metal-poor or metal-free stars in $z=3.1$
Ly$\alpha$ emitters\thanks{This work is based on data collected with the
Subaru Telescope operated by the National Astronomical Observatory of
Japan (program-ID: S07B-010 and S08B-046) and with the Very Large
Telescope operated by the European Southern Observatory, Chile
(program-ID: 081.A-0081).}}
\author[A.\ K.\ Inoue et al.]{\parbox[t]{\textwidth}{\vspace{-1cm}
A. K. Inoue$^{1}$\thanks{E-mail: akinoue@las.osaka-sandai.ac.jp},
K. Kousai$^{2}$, I. Iwata$^{3}$, Y. Matsuda$^{4}$, E. Nakamura$^{2}$, 
M. Horie$^{2}$, T. Hayashino$^{2}$, C. Tapken$^{5,6}$, M. Akiyama$^{7}$, 
S. Noll$^{8,9}$, T. Yamada$^{7}$, D. Burgarella$^{8}$ 
and Y. Nakamura$^{2}$}\\\\
$^{1}$College of General Education, Osaka Sangyo University, 
3-1-1, Nakagaito, Daito, Osaka 574-8530, Japan\\
$^{2}$Research Center for Neutrino Science, Graduate School of Science,
Tohoku University, Aramaki, Aoba-ku, Sendai 980-8578, Japan\\
$^{3}$Okayama Astrophysical Observatory, National Astronomical
Observatory of Japan, Honjo, Kamogata, Asakuchi, Okayama 719-0232,
Japan\\
$^{4}$Department of Physics, Durham University, South Road, Durham DH1
3LE\\
$^{5}$Max Plank Institute for Astronomy, 69117 Heidelberg, Germany\\
$^{6}$Astrophysikalisches Institut Potsdam, An der Sternwarte
16, D-14482 Potsdam, Germany\\
$^{7}$Astronomical Institute, Graduate School of Science, Tohoku
University, Aramaki, Aoba-ku, Sendai 980-8578, Japan\\
$^{8}$Laboratoire d’Astrophysique de Marseille, Observatoire
Astronomique de Marseille-Provence, 38 rue Fr\'ed\'eric
Joliot-Curie, 13388 Marseille Cedex 13, France\\
$^{9}$Institut f\"ur Astro- und Teilchenphysik, Universit\"at
Innsbruck, Technikerstr.\ 25/8, 6020 Innsbruck, Austria}
\begin{document}

\date{}

\pagerange{\pageref{firstpage}--\pageref{lastpage}} \pubyear{2009}

\maketitle

\label{firstpage}

\begin{abstract}
Deep narrowband (NB359) imaging with Subaru telescope by Iwata et
 al.\ have detected surprisingly strong Lyman continuum (LyC;
 $\sim900$ \AA\ in the rest-frame) from some Lyman $\alpha$ emitters
 (LAEs) at $z=3.1$. However, a possibility of redshift misidentification
 was not rejected with previous spectroscopy due to a narrow wavelength
 coverage. We here present results of new deep spectroscopy covering the
 observed 4,000--7,000 \AA\ with VLT/VIMOS and Subaru/FOCAS of 8 LAEs
 detected in NB359. All the eight objects have only one detectable
 emission line around 4,970 \AA\ which is most likely to be Ly$\alpha$
 at $z=3.1$, and thus, the objects are certainly LAEs at the redshift. 
 However, five of them show a $\sim0.''8$ spatial offset between the
 Ly$\alpha$ emission and the source detected in NB359. No indications of
 the redshifts of the NB359 sources are found although it is
 statistically difficult that all the five LAEs have a foreground object
 accounting for the NB359 flux. The rest three LAEs show no significant
 offset from the NB359 position. Therefore, we conclude that they are
 truly LyC emitting LAEs at $z=3.1$. We also examine the stellar
 population which simultaneously accounts for the strength of the LyC
 and the spectral slope of non-ionizing ultraviolet of the LAEs. We
 consider the latest statistics of Lyman limit systems to estimate the
 LyC optical depth in the IGM and an additional contribution of the
 bound-free LyC from photo-ionized nebulae to the LyC emissivity. As a
 result, we find that stellar populations with metallicity 
 $Z\geq1/50Z_\odot$ can explain the observed LyC strength
 only with a very top-heavy initial mass function (IMF; 
 $\langle m \rangle \sim 50 M_\odot$). However, the critical
 metallicity for such an IMF is expected to be much lower. A very young
 ($\sim1$ Myr) and massive ($\sim100$ $M_\odot$) extremely metal-poor
 ($Z\leq5\times10^{-4}Z_\odot$) or metal-free (so-called Population III)
 stellar population can reproduce the observed LyC strength. The
 required mass fraction of such `primordial' stellar population is
 $\sim1$--10\% in total stellar mass of the LAEs. We also present a
 possible evolutionary scenario of galaxies emitting strong LyC and
 implications of the primordial stars at $z\sim3$ for the metal
 enrichment in the intergalactic medium and for the ionizing background
 and reionization.
\end{abstract}


\section{Introduction}

The first generation of stars in the universe is the stellar
population without any metal elements, or so-called Population III
(Pop III) stars \citep[e.g.,][]{bro04}. This population is expected to
be as massive as $\sim100$ $M_\odot$ \citep[e.g.,][]{bro04} and is
thought to play an important role on the cosmic reionization at $z>6$
\citep[e.g.,][]{loe01}. How long did such metal-free star formation last
in the universe? Although the metal enrichment in the intergalactic
medium (IGM) is uncertain, if it is inefficient, metal-free haloes may
form until $z=5$ \citep{tre09}, $z=2.5$ \citep{tor07}, or even $z<2$
\citep{joh10}. The metal mixing process in the interstellar medium (ISM)
of a galaxy is also uncertain, but a significant amount of metal-free
gas may coexist with enriched gas for a few hundred Myr \citep{pan07}. 
If it is true, metal-free stars may continue to form in galaxies for
such a long time.

The He {\sc ii} $\lambda$1640 emission line is proposed as a signature
of Pop III stars because of their very hard ionizing spectrum expected
theoretically \citep{sch02,sch03}. \cite{nag08} made a wide and
deep survey of the emission line at $z\sim4$ and obtained an upper limit
on Pop III star formation rate density at the redshift. On the other hand,
the emission line has been found in a composite spectrum of Lyman break
galaxies (LBGs) at $z\sim3$ \citep{sha03,nol04}. \cite{jim06} 
claimed that it was a signature of the metal-free stellar population in
the LBGs. However, this He {\sc ii} line is broad (FWHM $\sim1500$ km
s$^{-1}$), and thus, it is probably caused by stellar winds from
Wolf-Rayet stars with `normal' metallicity \citep{sha03,nol04,bri08}.

Another signature of Pop III stars is a very large Ly$\alpha$ equivalent
width (EW) as $>240$ \AA\ \citep{mal02}. Some surveys of Lyman $\alpha$
emitters (LAEs) found galaxies with such a large Ly$\alpha$ EW
at $z=3.1$ \citep{nak10}, $z=4.5$ \citep{mal02} and $z=5.7$
\citep{shi06}. However, a clumpy dusty medium may boost the EW only
apparently \citep{neu91}. In addition, there is a large uncertainty on
the EW measurements because LAEs are so faint that we may not measure
their continuum level accurately enough.

The stellar population with metal mass fraction (or metallicity)
$Z<10^{-5}$, which is $<1/2000$ $Z_\odot$, is
classified as extremely metal-poor (EMP) stars \citep[e.g.,][]{bee05}
and may be the second generation of stars \citep{ume03}. Hundreds of
low-mass EMP stars have been found in the halo of the Galaxy 
\citep[e.g.,][]{bee05}. Because of their long lifetime, these stars are
survivors of the early stage of the formation of the Galaxy. On the 
other hand, it is expected that their high-mass counterpart existed in
the early days and died out until the current epoch \citep{tum06,kom07}.
Yet, there is no direct observational evidence of such massive EMP
stars at high-$z$.

Although there is no confident observational signature of primordial
massive stellar populations such as Pop III and massive EMP stars so
far, these populations should exist in the early universe. These stars
emit strong ionizing radiation which should have played an important
role in the cosmic reionization \citep[e.g.,][]{loe01} and have affected
the galaxy formation in the subsequent epoch \citep[e.g.,][]{sus04}.
Therefore, looking for these stellar populations is highly important to
understand the feedback process by the first generation of stars and
galaxies as well as to prove their existence.

Very recently, some candidates of galaxies at $z\ga7$ have been
discovered by the standard drop-out technique from the new data taken
with the WFC3/IR camera on the Hubble Space Telescope
\citep{bou09a,oes10,bun09,mcl09,yan09}. Interestingly, \cite{bou09b}
found that the ultraviolet (UV) colour of the $z\ga7$ candidate galaxies
was so blue that the galaxies may be composed of EMP or metal-free
stellar populations. However, we should confirm that the redshifts of
the galaxies are really $z\ga7$ with spectroscopy, which is too
difficult to do with current facilities.

\cite{ino09} (hereafter Paper I) has proposed a new method to find the
hypothetical stellar populations like massive EMP and metal-free stars at
$z\la4$. When Lyman continuum (LyC) emitted by stars escapes from
galaxies, the LyC emitted by photo-ionized nebulae around ionizing stars
may also escape. This bound-free nebular LyC has a peak just below the
Lyman limit, so that a spectral `bump' appears at the Lyman limit. We
call this `Lyman bump' or more precisely `Lyman limit bump'. The
strength of the Lyman bump depends on the hardness of the stellar LyC. 
Thus, more metal-poor galaxies will show stronger Lyman bump. On the
other hand, neutral hydrogen remains in the IGM even after the
completion of the reionization and it significantly absorbs the LyC. 
It makes difficult to find LyC or Lyman bump at $z>5$ \citep{ino08}.

Are there galaxies with a Lyman bump in the real universe? This paper
intends to show that the answer is yes. \cite{iwa09} (hereafter I09)
discovered 10 LAEs and 7 LBGs at $z\sim3$ with a significant leakage of
their LyC captured in a deep narrowband image taken with the
Subaru/Suprime-Cam (S-Cam). Some of the LAEs are indeed brighter in LyC
than in non-ionizing UV, strongly suggesting the presence of the Lyman
limit bump. However, there was a possibility that these objects are at
lower-$z$ because of a narrow wavelength coverage of previous
spectroscopy which only confirmed an emission line.

The rest of this paper is organized as follows. In section 2, we first
describe the sample of the possible LyC emitting LAEs discovered by
I09. In section 3, we present results of follow-up spectroscopy with
VLT/VIMOS and Subaru/FOCAS. In section 4, we thoroughly examine the
redshift of the LAEs (i.e. the reality of the detected LyC) with the
follow-up deep spectra and conclude that at least three LAEs are real
LyC emitters. In section 5, we compare the observed strength of the LyC
with the Lyman bump model proposed by Paper I and show that the LAEs are
likely to contain massive metal-free or EMP stars. In section 6, we
present a summary of our results, an evolutionary scenario of the LAEs,
and a few implications of the existence of primordial stars at $z\sim3$.

We adopt the AB magnitude system \citep{oke74} to describe object
magnitudes and colours in section 5. The standard flat $\Lambda$CDM
cosmology with $h=0.7$, $\Omega_{\rm M}=0.3$, and $\Omega_\Lambda=0.7$ 
is adopted if it is required.

\section{The sample of LAEs emitting strong Lyman continuum}

\begin{table*}
 \caption[]{Coordinates of the sample galaxies.}
 \setlength{\tabcolsep}{3pt}
 \footnotesize
 \begin{minipage}{\linewidth}
  \begin{tabular}{lccccccccccl}
   \hline
   & \multicolumn{2}{c}{NB359 position} 
   & \multicolumn{2}{c}{Offset of $R$} 
   & \multicolumn{2}{c}{Offset of NB497} 
   & NB359 $^a$ & $R$ $^a$ & \multicolumn{2}{c}{Spectroscopy} &\\
   Object & $\alpha$ (J2000)& $\delta$ (J2000) & $\Delta$ ($''$) 
   & PA ($^\circ$) & $\Delta$ ($''$) & PA ($^\circ$) & [AB] & [AB] 
   & Previous & Current & Remarks\\
   \hline
   a & 22:17:24.76 & $+$00:17:16.7 & 0.23 & 22 & 0.38 & 42 & 25.84 & 25.51 
    & FOCAS in 2005 & VIMOS/FOCAS & I09, LAB35 in M04\\ %(82931) \\
   b & 22:17:38.97 & $+$00:17:25.0 & 0.26 & 79 & 0.85 & 43 & 26.28 & 26.66 
    & FOCAS in 2005 & VIMOS/FOCAS & I09 \\ %(84007) \\
   c & 22:17:45.87 & $+$00:23:19.1 & 0.24 & 272 & 0.46 & 190 & 26.25 & 26.72 
    & FOCAS in 2005 & VIMOS & I09 \\ %(112839) \\
   d & 22:17:26.18 & $+$00:13:18.4 & 0.21 & 297 & 0.88 & 7 & 26.68 & 26.91 
    & DEIMOS in 2004 & VIMOS & I09 \\ %(64186) \\
   e & 22:17:16.64 & $+$00:23:07.4 & 0.26 & 76 & 1.06 & 32 & 26.73 & 26.24 
    & FOCAS in 2005 & VIMOS & I09 \\ %(112015) \\
   f & 22:17:08.04 & $+$00:19:31.7 & 0.25 & 341 & 0.34 & 7 & 26.88 & 26.52 
    & --- & VIMOS & \\ %(94460) \\
   g & 22:17:53.22 & $+$00:12:37.4 & 0.00 & --- & 0.20 & 257 & 26.84 & 26.84 
    & FOCAS in 2005 & VIMOS & \\ %(60925) \\
   h & 22:17:29.41 & $+$00:06:29.0 & 0.41 & 144 & 0.81 & 189 & 26.33 & 26.02 
    & --- & VIMOS & \\ %(31771) \\
   i & 22:17:12.74 & $+$00:28:55.4 & 0.16 & 4 & 0.12 & 320 & 26.26 & 24.59 
    & --- & VIMOS & broad-line AGN \\ %(144725) \\
   \hline
  \end{tabular}

  $^a$ Magnitudes within a $1.''2$ diameter aperture at NB359 or $R$
  position.
 \end{minipage}
\end{table*}%

We mainly deal with the sample of LAEs detected in their LyC by
I09. Since the detected LyC is too strong to be explained with a
standard stellar population model, I09 reported the detections but left
their nature as a mystery.

We started from a sample of LAEs and Ly$\alpha$ blobs
(LABs) selected through a Subaru/S-Cam narrowband NB497 imaging in the
SSA22 field \citep[][hereafter M04]{hay04,mat04}. The selection criteria
of the LAEs were NB497 $=20.0$--$26.2$ AB ($2.''0~\phi$, $>5\sigma$) and
the observed equivalent width of the emission line $>80$ \AA. Among
them, we had 125 galaxies with spectroscopic redshift $z_{\rm spec}>3.0$
measured with Subaru/FOCAS and Keck/DEIMOS (\citealt{mat05,mat06}, in
preparation; Yamada et al.\ in preparation). I09 performed a very deep
imaging with another narrowband filter NB359 with Subaru/S-Cam
(program-ID: S07B-010) in the same field. The NB359 filter exactly
captures LyC ($\simeq880$ \AA) for $z=3.1$ galaxies. In the NB359 image,
I09 found 10 objects identified with the spectroscopically
confirmed LAEs. The identification of NB359 (and $R$) sources with the
spectroscopic LAEs was done by the following procedure: (1) object
detections in NB359 and in $R$, (2) identification of each LAE with the
$R$ object nearest from the barycentre of NB497 intensity, and (3)
either identification of a NB359 object within $1.''4$ from the
barycentre of $R$ intensity or $>3\sigma$ detection at NB359 aperture
photometry with $1.''4$ diameter around the barycentre of $R$
intensity. Here we present results of follow-up deep spectroscopy with
VLT/VIMOS and Subaru/FOCAS for 5 out of the 10 LAEs in I09: the objects 
{\bf a}--{\bf e} in Table~1.

In addition, there are other $\sim30$ objects detected in the NB359
image and identified with the LAE candidates (i.e. not confirmed
spectroscopically yet) by the same procedure described above. Among
them, we also present results of deep spectroscopy with VIMOS for 4
objects: {\bf f}--{\bf i} in Table~1.\footnote{We recently found that
the spectrum of the object {\bf g} were taken with Subaru/FOCAS
previously (Yamada et al.\ in preparation).}

In Table~1, we summarise some properties of the sample LAEs: coordinates
in NB359 (LyC), offsets and position angles (PAs) of $R$ and NB497
(Ly$\alpha$) positions against the positions in NB359, and magnitudes in
NB359 and $R$. As found from the table, there are small offsets in the
three bands; the offsets between $R$ and NB359 are less than $0.''4$ and
those between NB359 and NB497 are $0.''1$--$1.''0$. The offsets of the
continuum emission traced by $R$ and NB359 of the LAE sample are small
and not significant relative to the positional uncertainty between the
two bands ($\sim0.''25$; I09). We should note that the $R$--NB359
offsets of the LBG sample reported in I09 is much larger and significant
($\sim0.''97$ on average). On the other hand, in some cases, the offsets
between NB359 and NB497 of the LAE sample are significant. This point
will be discussed in detail in \S3.2 and \S4.2.

\section{Spectroscopy of the LAEs}

\subsection{Previous medium-resolution spectroscopy}

We have several sets of previous spectroscopy with Subaru/FOCAS and
Keck/DEIMOS taken in 2003, 2004, and 2005 (\citealt{mat05,mat06}, in
preparation; Yamada et al.\ in preparation) and have confirmed that all
the 10 LAEs reported by I09 have a prominent emission line at around
4,970 \AA\ which was identified as Ly$\alpha$ at $z\simeq3.1$. DEIMOS
observations in 2004 and FOCAS observations in 2005 (\citealt{mat06}, 
in preparation; Yamada et al.\ in preparation) were made with a
relatively high spectral resolution ($R\sim2,000$) and rejected the
possibility that the emission line was the [O {\sc ii}] $\lambda$3727
doublet at $z=0.33$. However, the FOCAS observations in 2005 could not
exclude the possibility that the line was C {\sc iv} $\lambda$1549, He
{\sc ii} $\lambda$1640, C {\sc iii}] $\lambda$1909 or Mg {\sc ii}
$\lambda$2798 from an AGN at redshift $z=2.2$, 2.0, 1.6 or 0.78,
respectively, or H$\beta$ from a very faint galaxy at $z=0.022$ because
of a narrow wavelength coverage ($\Delta\lambda\sim$200 \AA). Since the
full width at half-maximum (FWHM) of the line was as narrow as 200--600
km s$^{-1}$, the objects would be a type 2 if they were AGNs. However,
some of the objects are spatially extended in the ${\rm NB497}-BV$ image
(i.e. line image), making the AGN interpretation unlikely. For example,
the object {\bf a} is identified as LAB by M04 and the object {\bf c}
has a FWHM of the line intensity profile of $1.''3$ against a $1.''0$
PSF (a significance of 3-$\sigma$).

\subsection{VLT/VIMOS low-resolution spectroscopy}

\begin{figure*}
 \begin{center}
  \includegraphics[width=16cm]{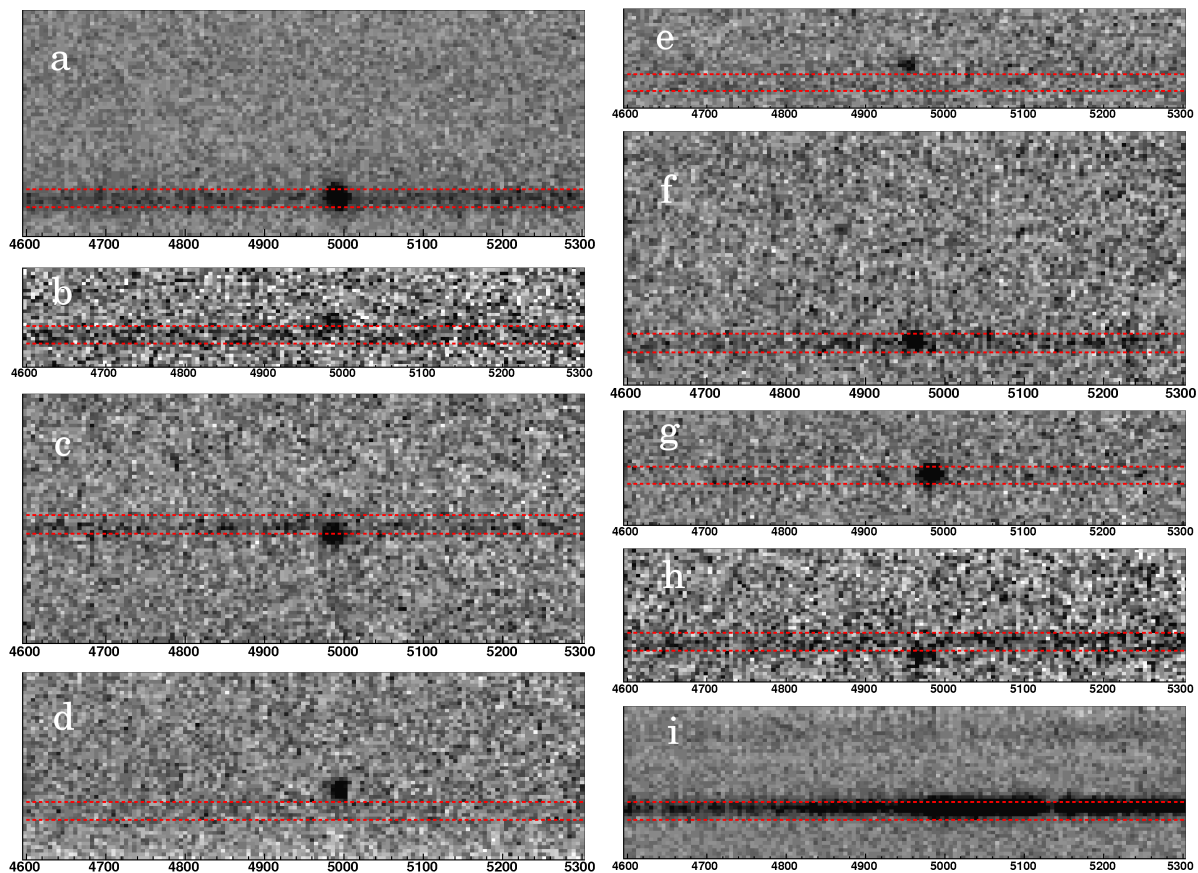}
 \end{center}
 \caption{Close-up around 5,000 \AA\ of the final two-dimensional spectra
 taken with VLT/VIMOS. The dotted lines in each panel show 5 pixels
 ($1.''0$) centring around the NB359 source position along the slitlet. 
 North is up.}
\end{figure*}

\begin{table}
 \caption[]{A summary of the VLT/VIMOS spectroscopy.}
 \setlength{\tabcolsep}{3pt}
 \footnotesize
 \begin{minipage}{\linewidth}
  \begin{tabular}{lcccc}
   \hline
   Object & Line offset ($''$)$^a$ & $z_{\rm Ly\alpha}$ & Decrement$^b$ \\
   \hline
   \multicolumn{4}{l}{Without line offset} \\
   a (con/emi) & 0.12 & 3.100 & $0.90\pm0.04$ \\
   f (con/emi) & 0.06 & 3.075 & $0.77\pm0.11$ \\
   g (con/emi) & 0.10 & 3.094 & $0.70\pm0.16$ \\
   composite$^c$ & --- & --- & $0.80\pm0.07$ \\
   \hline
   \multicolumn{4}{l}{With line offset} \\
   b (con) & 0.62 & --- & $1.18\pm0.29$ \\
   \phantom{b} (emi) & --- & 3.090 & $0.54\pm0.29$ \\
   c (con) & 0.51 & --- & $0.94\pm0.15$ \\
   \phantom{c} (emi) & --- & 3.095 & $1.38\pm0.31$ \\
   d (con) & 1.29 & --- & $1.49\pm0.68$ \\
   \phantom{d} (emi) & --- & 3.100 & $0.77\pm0.35$ \\
   e (con) & 1.00 & --- & $1.31\pm0.44$ \\
   \phantom{e} (emi) & --- & 3.065 & $0.69\pm0.24$ \\
   h (con) & 0.76 & --- & $0.73\pm0.11$ \\
   \phantom{h} (emi) & --- & 3.080 & $0.87\pm0.23$ \\
   composite$^c$ (con) & --- & --- & $1.12\pm0.15$ \\
   composite$^c$ (emi) & --- & --- & $0.86\pm0.13$ \\
   \hline
   \multicolumn{4}{l}{AGN} \\
   i (con/emi) & 0.25 & 3.11$^d$ & $0.72\pm0.03$ \\
   \hline
  \end{tabular}

  $^a$ Spatial offsets between the position detected in NB359 and the
  emission line.\\
  $^b$ Flux density ratios between 1050--1150 \AA\ and 1250--1350 \AA\ in
  the rest-frame of the Ly$\alpha$ redshift. Uncertainties are
  estimated from the quadratic combination of mean error of the
  continuum level and sky rms noise. The Galactic dust extinction has
  been corrected.\\
  $^c$ Average composite based on $z_{\rm Ly\alpha}$.\\
  $^d$ This redshift was determined from metal lines.
 \end{minipage}
\end{table}%

We performed a wide wavelength coverage (but low resolution)
follow-up spectroscopy for 5 (objects {\bf a}--{\bf e}) out of the 10
LAEs reported by I09 with VLT/VIMOS (program-ID: 081.A-0081). In
addition, we took spectra of 4 (object {\bf f}--{\bf i}) LAE candidates
detected in the NB359 image. Unlike the previous spectroscopy, we put
slitlets on the positions detected in the NB359 image in order to take
spectra of NB359 emitting sources. 

The observations were made with LR-Blue/OS-Blue grism 
($R=\lambda/\Delta \lambda\simeq180$ and dispersion 5.36 \AA/pix) and
$1.''0$ width slitlets on several dark nights during July to October
2008. The wavelength coverage of this setting is 3,700--6,800 \AA. The
plate scale is $0.''205$/pix. We took two masks (each has 4
quadrants) and the exposure time was 4 hours for each mask. There was an
overlap area of the two masks, so that the object {\bf a} was taken
twice. The observing conditions were good with typical seeing of
$0.''8$. The data reduction was done by a standard procedure with 
{\sc iraf}. The flux calibration was done with standard stars EG 274, G
158-100, and LTT 1020. Typical sky rms per pixel at around 5,000 \AA\ in
the resultant two dimensional spectra is about 0.02 $\mu$Jy.

Figure~1 shows close-up images around 5,000 \AA\ of the final
two-dimensional spectra of 9 objects observed with VLT/VIMOS. In each
panel, the area between the two dashed lines indicates 5 pixels ($=1.''0$)
centring around the position of the NB359 source. To find the NB359
source position, we used an {\sc iraf} task, {\sc geomatch}, between
the S-Cam image and the VIMOS pre-image. A typical uncertainty of this
procedure is found to be $\sim0.3$ pix ($=0.''06$). We also found a few
pix systematic offset between the pre-image and the spectral images and
corrected it based on the position of some bright objects detected in
both images (not the LAEs discussed in this paper). 

As found in Figure~1, a significant spatial offset between the emission
line and the NB359 position is evident in some objects. The measured
spatial offsets are summarised in Table~2 and consistent with those in
Table~1 measured in the two narrowband images. Relatively large offset
in the narrowband images of the object {\bf a} is probably due to the
extended Ly$\alpha$ emission classified as LAB. We categorise objects 
{\bf a}, {\bf f} and {\bf g} into the sample without line offset
(i.e. offset $<0.''2=1$ pix) and objects {\bf b}--{\bf e} and {\bf h}
into the sample with line offset based on the spectral image. The reason
why we rely on the spectral image is that we can measure the offset
between the continuum position and the pure emission line, whereas the
NB497 intensity traces both the line and the continuum. The last object 
{\bf i} is found to be a broad-line AGN from the spectrum.

Figures~2 and 3 show the one-dimensional spectra extracted at the NB359
source positions of the samples without and with the line offset,
respectively. Figure~4 shows the spectrum of the object {\bf i} which is
identified as a broad-line AGN. In Table~2, we also list the redshifts
measured assuming the emission line to be Ly$\alpha$ and the flux
density decrements of the continua blueward and redward from the
line. We measured the decrements both at the NB359 source position and
at the emission line position for the sample with line offset. We have
corrected the decrements for the dust extinction by the Milky Way based
on the extinction map by \cite{sch98}. Figure~5 shows average composite
spectra of both samples.

\begin{figure}
 \begin{center}
  \includegraphics[width=8cm]{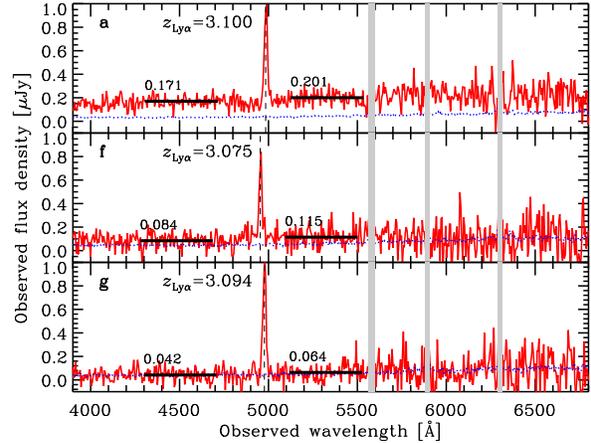}
 \end{center}
 \caption{VLT/VIMOS one-dimensional spectra of 3 objects without offset
 between the NB359 position and the emission line. The dotted curves are
 1-$\sigma$ sky noise spectra. The vertical dashed lines indicate the
 wavelengths of the emission lines detected in two-dimensional spectra
 shown in Figure~1. In each panel, we note the redshift if the emission
 line is Ly$\alpha$. The two thick horizontal lines with numbers in each
 panel show average continuum levels within 1050--1150 \AA\ and
 1250--1350 \AA\ in the rest-frame for the Ly$\alpha$ redshift. The
 values are not corrected for the Galactic extinction. The vertical
 shaded regions show the ranges affected by Earth's atmospheric lines.}
\end{figure}

\begin{figure}
 \begin{center}
  \includegraphics[width=8cm]{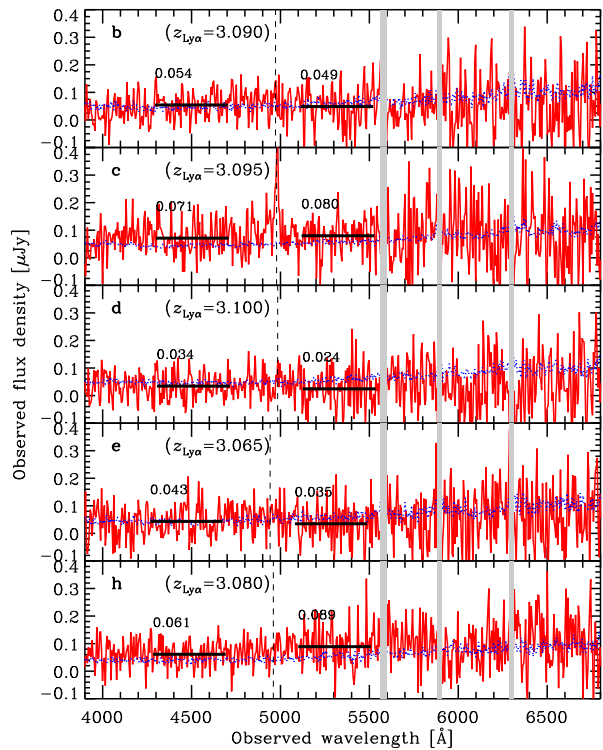}
 \end{center}
 \caption{Same as Figure~2 but for 5 objects with offset between the
 NB359 position and the emission line. The spectra are extracted at the
 NB359 positions.}
\end{figure}

\begin{figure}
 \begin{center}
  \includegraphics[width=8cm]{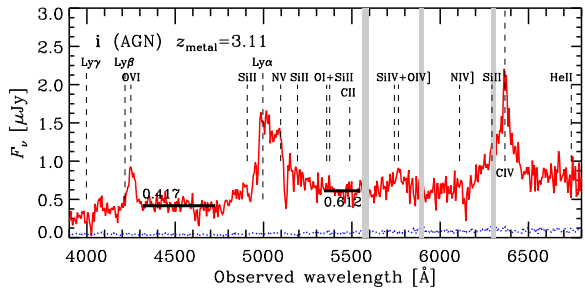}
 \end{center}
 \caption{Same as Figure~2 but for the object identified as a broad-line
 AGN. The redshift was determined by metal lines indicated by vertical
 dashed lines. The average continuum level of the longer wavelength is
 measured within 1300--1350 \AA\ to avoid an effect of the broad
 Ly$\alpha$ line.}
\end{figure}

\begin{figure}
 \begin{center}
  \includegraphics[width=8cm]{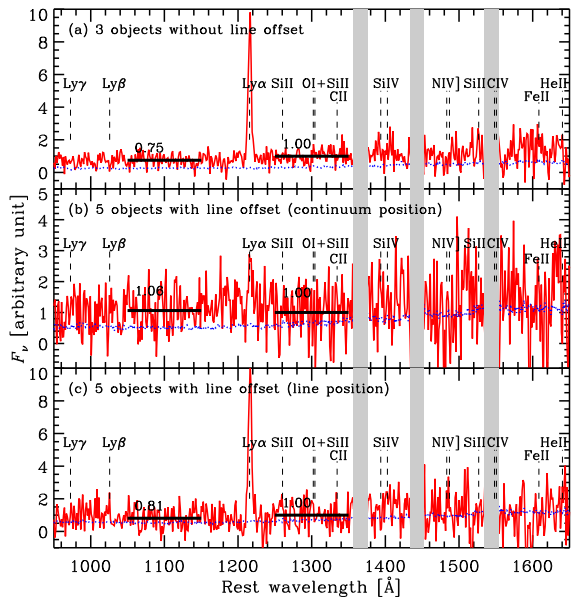}
 \end{center}
 \caption{Average composite spectra of (a) 3 objects without offset
 between the NB359 position and the emission line (objects {\bf a}, 
 {\bf f}, and {\bf g}), (b) 5 objects with offset (objects {\bf b}, 
 {\bf c}, {\bf d}, {\bf e}, and {\bf h}), and (c) the same 5 objects but
 at the emission line position. The dotted curves are 1-$\sigma$
 uncertainty estimated from sky noise spectra. We averaged the
 one-dimensional spectra in the rest-frame for the redshift assuming the
 emission line to be Ly$\alpha$. The thick horizontal lines with numbers
 indicate mean flux densities within 1050--1150 \AA\ and 1250--1350
 \AA. At the latter wavelength range, we normalised the spectra before
 making an average. The values are not corrected for the Galactic
 extinction. The vertical dashed lines indicate wavelengths of
 emission/absorption lines. The vertical shaded regions show the ranges
 affected by Earth's atmospheric lines.}
\end{figure}

\subsection{Subaru/FOCAS low-resolution spectroscopy}

We have spectra of two objects {\bf a} and {\bf b} taken with
Subaru/FOCAS (program-ID: S08B-046). The observation was done with the
300B grism (dispersion: 1.34 \AA/pix) and the slit width of $0.''8$
during the dark night on 22 September 2008. The wavelength coverage of
this observation is 4,000--7,200 \AA. The resultant spectrum of the
object {\bf a} is very consistent with those in Figure~2 but has a
higher spectral resolution. On the other hand, we could 
not find any emission line in the spectrum of the object {\bf b}. This
is because the position angle of the FOCAS slitlet (117$^\circ$) was far
from that of the emission line from the NB359 position (43$^\circ$; see
Table~1). The position angle of the VIMOS slitlet was 0$^\circ$
(north-south) and we did capture a part of the emission line as shown in
Figure~1.

\section{Reality of the Lyman continuum}

In this section, we discuss whether NB359 fluxes detected from the
sample LAEs are truly LyC or not. Namely, we discuss whether the
redshifts of the continuum sources are $z\simeq3.1$ or not.
As found in Figure~1, some objects show a spatial offset of the
emission line at around 4,970 \AA. We deal with the two samples with or
without the line offset separately.

\subsection{Sample without line offset}

\subsubsection{Line identification}

From the objects {\bf a}, {\bf f}, and {\bf g}, we clearly detect an
emission line at around 4,970 \AA\ and no significant spatial offset
from the position detected in NB359. Therefore, the emission line is
highly likely to come from the continuum source detected in NB359 (see
\S4.1.2 about the probability of foreground contamination). Finally, we
conclude that the emission line is Ly$\alpha$ at $z\simeq3.1$ and
the NB359 flux is truly LyC, based on the following considerations.

We could not detect any significant emission/absorption lines except for
the 4,970 \AA\ line from the 3 galaxies shown in Figure~2. Upper limits
(3-$\sigma$) on the line flux ratio relative to the 4,970 \AA\ line are 
$<0.10$--0.18 at 4,000 \AA, $<0.10$--0.18 at 4,500 \AA,
$<0.11$--0.20 at 5,000 \AA, $<0.15$--0.24 at 5,500 \AA, 
$<0.20$--0.35 at 6,000 \AA, and $<0.24$--0.42 at 6,500 \AA. 
These limits are estimated from a comparison between a typical sky rms
noise for 5-pix in spatial direction at a wavelength and the 4,970 \AA\
line flux. We have assumed that the postulated emission line has the
same width as the 4,970 \AA\ line. This assumption is justified if the
lines are narrower than about 1,600 km s$^{-1}$ (i.e. $R\simeq180$). 
These 3-$\sigma$ upper limits of the line flux ratio indicate that we
can exclude the existence of other emission lines even if the lines are
a factor of 0.1 to 0.5 weaker than the 4,970 \AA\ emission line.

Let us identify the 4,970 \AA\ line. We consider 7 possible
identifications of the line. If the 4,970 \AA\ line was C {\sc iv} and
the galaxies were AGNs at $z\simeq2.2$, we should have Ly$\alpha$ around
3,900 \AA. Based on the QSO average spectrum by \cite{fra91}, the flux
ratio of Ly$\alpha$ to C {\sc iv} is 1.6. Thus, we should detect the
Ly$\alpha$ in our spectra with a high significance. However, we could
not. We could not detect He {\sc ii} and C {\sc iii}] which have flux
ratios relative to C {\sc iv} of 0.3 and 0.5, respectively
\citep{fra91,hec95,kro99}, either. If the 4,970 \AA\ line was He 
{\sc ii} or C {\sc iii}] and the galaxies were AGNs at $z\simeq2.0$ or
1.6, we could detect C {\sc iv} which is stronger than the two lines
\citep{fra91,hec95,kro99} at around 4,690 \AA\ or 4,030 \AA. In the Mg
{\sc ii} case, we could detect a stronger [O {\sc ii}] emission line
from narrow-line regions at around 6,630 \AA\ \citep{kro99}. In the 
[O {\sc ii}] or the H$\beta$ cases, we could detect stronger 
[O {\sc iii}] emission lines at around 6595/6659 \AA\ or 5,070/5,120
\AA, respectively. However, we could not at all. The [O {\sc ii}]
interpretation had been already ruled out by the previous
medium-resolution spectra for the objects {\bf a}--{\bf e} and {\bf g}.
In the case of Ly$\alpha$ from star-forming galaxies at $z=3.1$, we do
not expect any other stronger emission lines within the wavelength
coverage. Therefore, the single emission line in the wide wavelength
coverage strongly suggests that the line is Ly$\alpha$ and the objects
lie at $z\simeq3.1$.

If the galaxies are at $z=3.1$, we expect Ly$\alpha$ decrement to be
$0.75\pm0.03$ on average at around 1,100 \AA\ in rest-frame of the
galaxies \citep{ino08,fau08} (Note that we are looking at Ly$\alpha$
forest at $z=2.7$ in this wavelength.). Let us see the decrement in the
spectra of the 3 objects ({\bf a}, {\bf f} and {\bf g}) and the AGN
(object {\bf i}). Comparing the flux density within the rest-frame
1,050--1,150 \AA\ to that within 1,250--1,350 \AA\ with the assumptions
of the 4,970 \AA\ line to be Ly$\alpha$ and a flat intrinsic spectrum
between 1,100 \AA\ and 1,300 \AA\ in the rest-frame, we find decrements
of 0.7--0.9 after correcting for the Galactic extinction as summarised
in Table~2. Although the decrement of the object {\bf a} is somewhat
smaller than the expected, other three decrements are very consistent
with the Ly$\alpha$ forest measurement, again supporting our 
interpretation that they are $z\simeq3.1$ objects.

In summary, the 3 objects without line offset have a strong and narrow
emission line around 4,970 \AA\ but do not have any other detectable
emission/absorption lines. Their continuum shows a small break below the
emission line and the amount of the decrement is consistent with that
expected from Ly$\alpha$ forest if they are at $z\simeq3.1$. Therefore,
we conclude that the emission line is redshifted Ly$\alpha$ and the
objects are truly LAEs at $z\simeq3.1$.

\subsubsection{Probability of foreground contamination}

Even when the emission line around 4,970 \AA\ is the $z\simeq3.1$
Ly$\alpha$, the LAEs may have a foreground object on the line of sight
towards them and the flux detected in NB359 may come from the low-$z$
interloper. This possibility is extensively discussed in \cite{van10} 
\citep[see also][and I09]{sia07}. Let us estimate the probability to
have a contamination in front of the 3 LAEs discussed in this subsection
according to \cite{van10}.

The 3 LAEs without line offset ({\bf a}, {\bf f} and {\bf g}) are
detected in NB359 with a magnitude of 26--27 AB (see Table~1) and the
spatial offset between the Ly$\alpha$ emission line and the continuum is
less than $0.''2$ ($=1$ pix) (see Table~2). Based on the $U$-band number
count by \cite{non09} of 130,200 deg$^{-2}$ in this magnitude range, we
expect only 0.126\% of chance coincidence of an object within $0.''2$
around another object.

The object {\bf a} is found from 10 objects reported by I09 among 125
spectroscopic LAE sample. The 10 objects may include foreground
contamination, but the probability that one object picked up from
the 10 objects has a foreground within $0.''2$ is only 1.6\%\footnote{If 
each object has a foreground with a probability $p$, the probability to
have $k$ foregrounds among $n$ objects is 
$\left(n \atop k\right)p^k(1-p)^{n-k}$ \citep{van10}. If we make a
subsample of $m(\ge k)$ objects including the $k$ foregrounds and we
randomly select $i$ objects from the $m$ objects, the probability to
have $j(\le i, k)$ foregrounds among the $i$ objects is 
$\left(k\atop j\right)\left(m-k\atop i-j\right)/\left(m\atop i\right)$.
Since $k$ can vary from $j$ to $m-i+j$, the total probability to have $j$
foregrounds among $i$ objects becomes 
$\sum_{k=j}^{m-i+j} \left(n \atop k\right)p^k(1-p)^{n-k} 
\left(k\atop j\right)\left(m-k\atop i-j\right)/\left(m\atop i\right)$.}.
The objects {\bf f} and {\bf g} are found from $\sim30$ objects detected
in NB359 among $\sim800$ LAE candidates. If we take two objects from the
$\sim30$ sample, the probability that both two have a foreground within
$0.''2$ is only 0.1\% and the probability that one has a foreground is
still 6.5\%. Therefore, the three objects {\bf a}, {\bf f} and {\bf g}
are unlikely to be contaminated by foreground objects.

\subsubsection{Spectroscopic properties}

We here note a few more
measurements from the spectra. Table~3 is a summary of the derived
properties. The FWHMs of Ly$\alpha$ are smaller than the velocity
resolution of our observations ($\simeq1,600$ km s$^{-1}$). Thus, the
line width is unresolved. The EWs of Ly$\alpha$ in the rest-frame are
20--80 \AA\ in individual spectra and 43 \AA\ in the average spectrum,
which are not very large as expected for Pop III stars 
\citep[e.g.,][]{mal02,sch02}. If LyC escape fraction is relatively
large, however, the expected EW of Ly$\alpha$ decreases. The EW is
further reduced by neutral hydrogen in the ISM and in the IGM
because of resonant scattering. Therefore, the measured Ly$\alpha$ EWs
do not reject the existence of Pop III stars in the three LAEs.

The He {\sc ii} $\lambda$1640 line is not detected and its 3-$\sigma$
upper limits relative to Ly$\alpha$ flux ($F_{\rm HeII}/F_{\rm Ly\alpha}$) 
are $<0.2$--0.4, where we have assumed that the He {\sc ii} line has the
same width as Ly$\alpha$. According to \cite{sch03}, the intrinsic
$F_{\rm HeII}/F_{\rm Ly\alpha}<0.02$ is expected for a Pop III star
cluster. Since only Ly$\alpha$ photons will be scattered resonantly in
the surrounding ISM and IGM, the observed $F_{\rm HeII}/F_{\rm Ly\alpha}$ 
can be enhanced significantly. On the other hand, the He {\sc ii} line
is detectable only in a short time-scale ($<$ a few Myr) from the Pop
III star formation \citep{sch02}. In any case, the obtained upper limits
on $F_{\rm HeII}/F_{\rm Ly\alpha}$ are not strong enough to reject the
presence of Pop III stars.

We could not detect any other emission/absorption lines in the spectra. 
The 3-$\sigma$ flux upper limits of emission lines are 0.1--0.4 relative
to Ly$\alpha$ depending on the wavelength (\S4.1.1). In this respect,
the three LAEs presented here are different from the Lynx arc at
$z=3.357$ and a peculiar LAE at $z=5.563$ emitting strong metal emission
lines such as N {\sc iv}] $\lambda1486$, C {\sc iv} $\lambda1549$, O
{\sc iii}] $\lambda1666$ and C {\sc iii}] $\lambda1909$ as well as
Ly$\alpha$ \citep{fos03,rai10,van10b}. The observed fluxes of 
N {\sc iv}] and C {\sc iv}, which are in the wavelength coverage of our
spectra, relative to Ly$\alpha$ are 0.1--0.3, which could be marginally
detected in our spectra if there were. These two peculiar galaxies are
likely to have very hot ($\sim10^5$ K) exciting stars
\citep{fos03,rai10} or small AGN \citep{van10b} and are as low
metallicity as $Z=1/20Z_\odot$. On the other hand, our three
LAEs may have massive EMP or Pop III stars, whose effective temperatures
are also $\sim10^5$ K, as shown later (\S5). Nevertheless, we could not
detect such metal emission lines. This may imply that the metallicity of
our three LAEs is much lower than $Z=1/20Z_\odot$ which is consistent
with the presence of EMP or Pop III stars.

\begin{table}
 \caption[]{Spectroscopic properties of the LAEs without line offset.}
 \setlength{\tabcolsep}{3pt}
 \footnotesize
 \begin{minipage}{\linewidth}
  \begin{tabular}{lccccc}
   \hline
   Object & $z_{\rm Ly\alpha}$ & FWHM$^a$ (km s$^{-1}$) 
   & EW$_0$$^b$ (\AA) & $F_{\rm HeII}/F_{\rm Ly\alpha}$$^c$ \\
   \hline
   a & 3.100 & 1190 & $20.4\pm1.1$ & $<0.24$ \\
   f & 3.075 & 1070 & $29.3\pm3.4$ & $<0.40$ \\
   g & 3.094 & 1130 & $78.5\pm12.3$ & $<0.25$ \\
   \hline
   composite$^d$ & --- & 1140 & $43.2\pm2.9$ & $<0.21$ \\
   \hline
  \end{tabular}

  $^a$ Full width at half-maximum of Ly$\alpha$.\\
  $^b$ Rest-frame equivalent widths of Ly$\alpha$. Uncertainties include
  only statistical errors in estimating the line flux based on sky rms
  noise.\\
  $^c$ 3-$\sigma$ upper limits on the He {\sc ii} line flux relative to
  Ly$\alpha$ estimated from sky rms noise. The line width is assumed to
  be the same as that of Ly$\alpha$.\\
  $^d$ Average composite.
 \end{minipage}
\end{table}%

\subsection{Sample with line offset}

\subsubsection{Line identification}

We detected a prominent emission line at 4,970 \AA\ around the
objects {\bf b}--{\bf e}, and {\bf h} as shown in Figure~1. However, the
line offsets from the NB359 source position by $0.''5$--$1.''3$,  which
correspond to 3.8--9.9 kpc (proper) if the objects lie at $z=3.1$. In
the emission line positions, we find no other emission/absorption
lines. After a very similar discussion to \S4.1.1, we conclude that the
emission line is likely to be Ly$\alpha$ at $z\simeq3.1$. In the NB359
positions, we do not find any significant emission/absorption lines,
either.

We measured Ly$\alpha$ decrements in both NB359 and emission line 
positions of each object, assuming the emission line to be Ly$\alpha$,
and the results are summarised in Table~2. We detect a significant
($>2$-$\sigma$) decrement at the continuum position of the object 
{\bf h} and marginal (1--2-$\sigma$) ones at the line position of the
objects {\bf b} and {\bf e}. Other cases are not conclusive because of
not-significant continua although some cases may imply no decrements. 
The decrements in the composite spectra are not conclusive, either, 
although that at the NB359 positions exceeds unity, indicating
contaminations of foreground objects.

In summary, the emission lines detected in our spectra are likely to be
Ly$\alpha$, and thus, these objects are LAEs at $z\simeq3.1$. However,
almost no information about redshift of the NB359 continuum sources was
extracted from our spectra. These sources may be physically associated
with the close LAEs and their redshift may be $z\simeq3.1$. Or these are
just foreground objects apparently close to the LAEs.

\subsubsection{Probability of foreground contamination}

As discussed in \S4.1.2, there is a possibility that a foreground
interloper accounts for the NB359 flux (and other optical band fluxes).
The 5 objects we are discussing here show an offset from Ly$\alpha$ line
by $0.''8$ on average (see Table~2). Let us estimate the probability of
such foreground contamination according to \cite{van10} as done in
\S4.1.2.

The observed NB359 magnitudes of the 5 objects ({\bf b}--{\bf e} and
{\bf h}) are 26--27 AB same as the 3 LAEs without line offset. Then, the
probability to have an object within $0.''8$ around another object is
2.02\% based on the spatial density of $U=26$--27 AB sources reported by
\cite{non09}. 

The objects {\bf b}--{\bf e} are selected from 10 LAEs which were 
found from 125 spectroscopic LAEs and reported by I09. The probability
that all the four taken from the 10 sample have a foreground within
$0.''8$ is only 0.8\%, but the probabilities that 3, 2, 1, or 0 objects
have a foreground are 5.7\%, 20.6\%, 39.8\%, or 33.2\%. The object
{\bf h} are found from $\sim30$ objects detected in NB359 among
$\sim800$ LAE candidates. The probability that the object is
contaminated by a foreground within $0.''8$ is 53.8\%.
Therefore, a few of the five objects are likely to be contaminated by a
foreground. However, {\it it is very difficult to conclude that all the
five are contaminated.} 

As found in Table~2, the objects {\bf d} and {\bf e} have a relatively
large offset ($>1.''0$) and no indication of Ly$\alpha$ decrement at the
NB359 positions. It may suggest that these are contaminations. On the
other hand, the object {\bf c} has the smallest offset and a possible
indication of decrement, may suggesting that it is a real LyC
emitter. However, it is impossible to conclude which objects are
contaminations using the current data.

\subsubsection{Possible causes of Ly$\alpha$ line offset}

We can still expect that a few objects among the 5 are associated with
LAEs at $z\simeq3.1$ and the NB359 traces the LyC according to the
statistical argument obtained in \S4.2.2. In other words, 
the offset between the Ly$\alpha$ emission line and the continuum could
be real in a few cases. For example, it is possible that a galaxy is
composed of multiple substructures and every component emits both the
continuum and the emission line. However, the line flux relative to the
continuum (i.e. EW) are different from each other. In fact, such a
situation is observed in local galaxies emitting Ly$\alpha$
\citep{ost09}. Then, if we observe the galaxy as a single object with an
insufficient spatial resolution, the barycentre of the emission line
intensity and of the continuum will offset. 

Another possibility is that Ly$\alpha$ emission comes from `cold
accretion' into galaxies \citep{dij09,goe10,fau10}. The `cold accretion
(or stream)' is thought to be the main mode to feed gas to galaxies
\citep{dek09}, and its temperature is $\sim10^4$ K, so that it emits
Ly$\alpha$. Galaxies have multiple streams and Ly$\alpha$ emission from
the streams is spatially extended \citep{dij09,goe10,fau10}. If we
observe such an extended Ly$\alpha$ emission with insufficient spatial
resolution, the barycentre of the emission may offset from the stellar
component.

Therefore, the Ly$\alpha$ offset may not be a surprising event for LAEs
in general. However, we have no further clue to resolve which is a real
$z\simeq3.1$ object for the moment.

\section{Rest UV two-colour diagram}

In this section, we try to interpret atypical UV colours of the LAEs
(and also LBGs) detected in LyC. Figure~6 shows a two-colour diagram in
the rest-frame UV of the LAEs and LBGs. We consider that the 3 LAEs
without line offset (filled circles) are emitting LyC and a
few of the 5 LAEs with line offset (open circles) are possibly emitting
LyC. The 7 LBGs reported by I09 are also shown by filled
squares. Additionally, we show the AGN (object {\bf i}) found in this
study. The colours are measured at the barycentre in $R$ with a circular
aperture whose diameter is chosen from $1.''2$ to $2.''6$ for each
object so as to include its total flux as done in I09.\footnote{There
was a revision of zero-points in $V$, $R$, and $i'$ after publication of
I09 and the resultant shifts in the two colour diagram as follows:
$\Delta({\rm NB359}-R)=+0.1$ and $\Delta(V-i')=+0.1$. Although the
zero-point revision is almost settled, we conservatively apply additional
systematic uncertainties of 0.1 mag to the error-bars of the observed
points by the quadratic combination with statistical uncertainties in
photometry.} Note that there are differences from Table~1 because of the
different method of measurements.

\begin{figure*}
 \begin{center}
  \includegraphics[width=12cm]{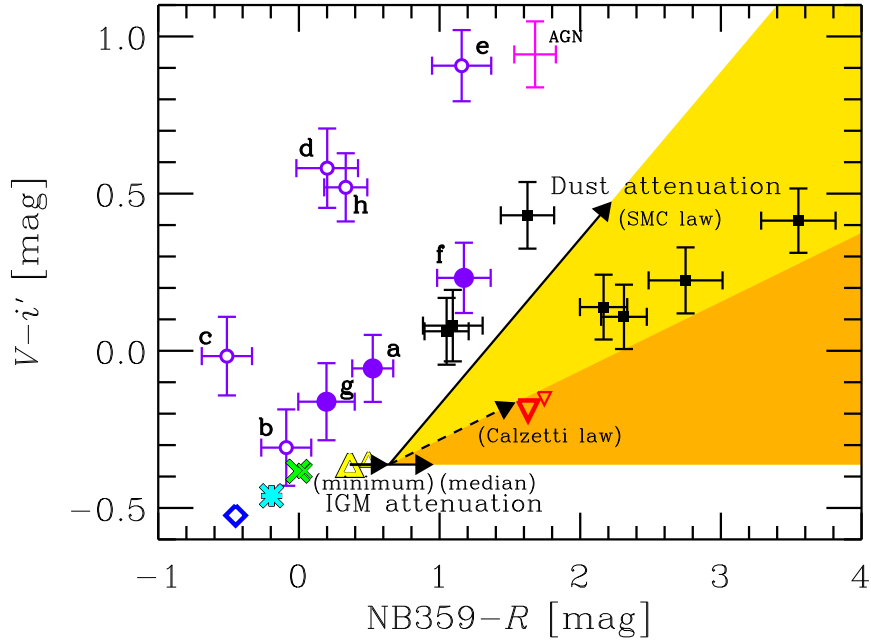}
 \end{center}
 \caption{Rest UV two-colour diagram for $z\simeq3.1$ LyC emitting
 galaxies. The vertical axis, $V-i'$, indicates non-ionizing UV
 spectral slope, and the horizontal axis, NB359$-R$, indicates
 LyC-to-UV flux density ratio. The filled circles with errorbars are the
 3 LAEs without line offset and the open circles with errorbars are the
 5 LAEs with line offset. The squares with errorbars are LBGs reported
 in Iwata et al.~(2009). The plus mark with errorbars is an AGN emitting
 LyC found in this study. The intrinsic colours of stellar population
 models are shown by diamond (the model D in Table 4), asterisk (C),
 x-marks (large: B1; small: B2), triangles (large: A1; small: A2) and
 inverse triangles (large: Ac1 with 100 Myr; small: Ac2 with 100 Myr). 
 The two horizontal arrows show IGM minimum and median attenuations for
 $z=3.1$. The two diagonal arrows show dust attenuations with
 $E(B-V)=0.1$ for the SMC law (solid arrow) and the Calzetti law (dashed
 arrow). The shaded regions indicate the regions explained by the
 stellar population model A1 with a combination of IGM and dust
 attenuations.}
\end{figure*}

\subsection{Stellar population models}

\begin{table*}
 \caption[]{Models of stellar population and SED.}
 \setlength{\tabcolsep}{3pt}
 \footnotesize
 \begin{minipage}{\linewidth}
  \begin{tabular}{lcccccccc}
   \hline
   Model & $Z/Z_\odot$ & $p$ & $m_{\rm up}$ & $m_{\rm low}$ 
   & $\langle m \rangle$ & SF history & age & SED reference\\
   \hline
   A1 & 1/50 & $-2.35$ & $100 M_\odot$ & $1 M_\odot$ & $3.1 M_\odot$
	& Instantaneous & 1 Myr & {\sc starburst}99 (v.5.1)\\
   A2 & 1/5 & $-2.35$ & $100 M_\odot$ & $1 M_\odot$ & $3.1 M_\odot$
	& Instantaneous & 1 Myr & {\sc starburst}99 (v.5.1)\\
   Ac1 & 1/50 & $-2.35$ & $100 M_\odot$ & $1 M_\odot$ & $3.1 M_\odot$
	& Constant & 10, 100, 1000 Myr & {\sc starburst}99 (v.5.1)\\
   Ac2 & 1/5 & $-2.35$ & $100 M_\odot$ & $1 M_\odot$ & $3.1 M_\odot$
	& Constant & 10, 100, 1000 Myr & {\sc starburst}99 (v.5.1)\\
   B1 & 1/50 & $-0.1$ & $100 M_\odot$ & $1 M_\odot$ & $48 M_\odot$
	& Instantaneous & 1 Myr & {\sc starburst}99 (v.5.1)\\
   B2 & 1/5 & $-0.1$ & $100 M_\odot$ & $1 M_\odot$ & $48 M_\odot$
	& Instantaneous & 1 Myr & {\sc starburst}99 (v.5.1)\\
   C & 1/2000 & $-2.35$ & $500 M_\odot$ & $50 M_\odot$ & $111 M_\odot$
	& Instantaneous & 1 Myr & Schaerer (2003)\\
   D & 0 & $-2.35$ & $500 M_\odot$ & $50 M_\odot$ & $111 M_\odot$
        & Instantaneous & 1 Myr & Schaerer (2003)\\
   \hline
  \end{tabular}
 \end{minipage}
\end{table*}%

In order to interpret the UV colours of the LyC emitting LAEs, we
construct models of spectral energy distributions (SEDs) of galaxies.
The SED of pure stellar populations depends on metallicity, initial mass
function (IMF), star formation history, and age. Since we are dealing
with LAEs (and LBGs), we consider sub-solar metallicities:
$Z=1/5\,Z_\odot$ and $1/50\,Z_\odot$. In addition, we consider
two more metallicities: $Z=1/2000\,Z_\odot$ (EMP) and 0
(metal-free). SEDs for the former two metallicities are generated by
the population synthesis code {\sc starburst99} version 5.1
\citep{lei99}. Those for the latter two are taken from
\cite{sch02,sch03}. The IMF is basically assumed to be Salpeter's one
\citep{sal55}: $\phi(m)dm\propto m^p dm$ with $p=-2.35$. Additionally,
we consider an extremely top-heavy case with $p=-0.1$. The mass range of
the IMF is assumed to be $m_{\rm up}=100$ $M_\odot$ and $m_{\rm low}=1$ 
$M_\odot$ for $Z=1/5\,Z_\odot$ and $1/50\,Z_\odot$ cases but to be 
$m_{\rm up}=500$ $M_\odot$ and $m_{\rm low}=50$ $M_\odot$ for the EMP
and metal-free cases. The average stellar masses are 3.1 $M_\odot$, 
48 $M_\odot$, or 111 $M_\odot$ depending on the IMF. We consider age of
1 Myr after an instantaneous star formation or 10 Myr to 1 Gyr constant
star formation. Table~4 is a summary of the models and their parameters.

In Figure~6, the intrinsic (i.e. no dust and IGM attenuations) colours
of the stellar population models are shown by diamond (metal-free: the
model D), asterisk (EMP: the model C), x-marks (extremely top-heavy IMF
but normal sub-solar metallicity: the models B1 and B2), triangles
(normal IMF and sub-solar metallicity: the models A1 and A2), and
inverse triangles (normal IMF and sub-solar metallicity but constant
star formation: the models Ac1 and Ac2). NB359$-R$ colours of the 3 LAEs
without line offset (large filled circles) and LBGs (small squares)
seem to be explained by a model with normal IMF and metallicity (models
A and Ac). However, the IGM attenuation makes NB359$-R$ redder as
described in the next subsection.

\subsection{IGM attenuation}

Since the IGM attenuation, especially for the LyC, has a stochastic
nature, we estimate the amount based on a Monte-Carlo simulation by
\cite{ino08} (hereafter II08). These authors assumed an empirical
distribution function of the intervening absorbers in redshift, column
density, and Doppler parameter spaces which was derived from the latest
observational statistics in that time. Their simulation reproduces the
Ly$\alpha$ depressions at $z=0$--6 very well. 

The LyC opacity is mainly determined by the Lyman limit systems (LLSs;
clouds with $\log_{\rm 10} (N_{\rm HI}/{\rm cm^{-2}})\geq17.2$), we
compare the number density evolution adopted in II08 with new survey
results after the paper in Figure~7. We find that observational data are
not converged yet; \cite{pro10} obtained smaller number of LLSs at $z<4$
than those by \cite{per05} and \cite{son10}. We also find that the
simulation by II08 adopted a number density tracing upper bounds of
observations. In order to match the new observations, we have reduced
the LLS number density, with almost keeping the number densities of
Ly$\alpha$ forest and damped Ly$\alpha$ systems\footnote{This is 
realized only if we change three parameters among ten in the empirical
distribution function of absorbers. The change is 
$({\cal A},\beta_1,\beta_2)=(400,1.6,1.3)$ to $(500,1.7,1.2)$,
where ${\cal A}$ is the normalisation of the number of absorbers, and
$\beta_1$ and $\beta_2$ are the indices of the double power-law for the
column density distribution. Note that we should keep the number density
of Ly$\alpha$ forest because the Ly$\alpha$ decrements by II08
excellently agree with observations.}. The updated number density
evolution is shown by diamonds in Figure~7. The new number density of
LLSs at $z=2.9$ which attenuate the NB359 flux of $z=3.09$ sources is
about 60\% of the previous one.

\begin{figure}
 \begin{center}
  \includegraphics[width=6cm]{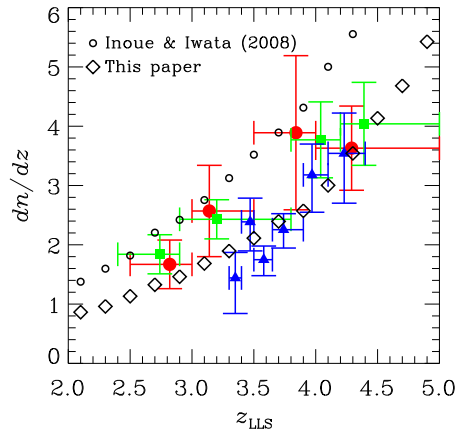}
 \end{center}
 \caption{The number density evolution of Lyman limit systems 
 which have $\log_{\rm 10} (N_{\rm HI}/{\rm cm^{-2}})\geq17.2$.
 Observational data are the filled circles (P{\'e}roux et al.~2005),
 squares (Songaila \& Cowie 2010), and triangles (Prochaska et al.~2010;
 a factor of 1.1 larger than their Table~4 which are the density for 
 $\log_{\rm 10} (N_{\rm HI}/{\rm cm^{-2}})\geq17.5$). Monte-Carlo
 simulations are shown by the open circles (Inoue \& Iwata 2008) and the
 diamonds (updated version).}
\end{figure}

\begin{figure}
 \begin{center}
  \includegraphics[width=6cm]{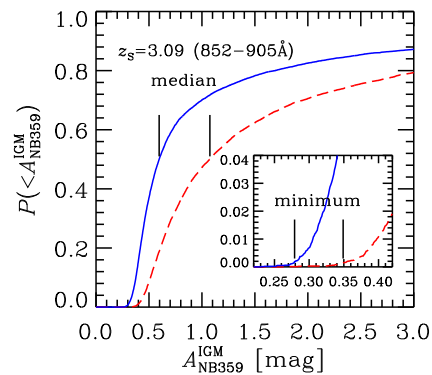}
 \end{center}
 \caption{The cumulative probability to have an IGM attenuation in NB359
 for $z=3.09$ sources smaller than that in the horizontal axis, based on
 our Monte-Carlo simulations: the dashed line (Inoue \& Iwata 2008) and
 the solid line (updated version). The number of realizations of lines
 of sight is 10,000. The narrowband filter NB359 captures 852--905 \AA\
 in the rest-frame of the sources. The inset shows a close-up of the
 smallest attenuation.}
\end{figure}

Figure~8 shows the cumulative probability of the IGM attenuation through
the NB359 for sources at $z=3.09$ which is the typical redshift of our
LAEs (and LBGs). To estimate the attenuation, we have assumed a constant
source spectrum in $f_\nu$ unit. The effect of the different shape of
the spectrum is negligible because the wavelength coverage in the NB359
is narrow. As found in the figure, the updated version (solid line)
expects much smaller attenuation than the original II08 (dashed line).
The median attenuation is reduced from 1.08 mag to 0.60 mag. Thus, the
LLS number density has a significant impact on the expected IGM
attenuation. We also find that the attenuation of the smallest 0.15\%
(e.g., smaller 3-$\sigma$ excess for Gaussian) is 0.35 mag (II08) or
0.28 mag (updated). These attenuations are realized on a line of sight
without LLS at $z\simeq3$ and produced by numerous and unavoidable
Ly$\alpha$ forest. Because the update in the number density of absorbers
is small for Ly$\alpha$ forest, the reduction of the smallest
attenuations is small. We adopt the attenuation at the cumulative
probability of 0.15\% as the minimum IGM attenuation in the following.

In Figure~6, we show the minimum and median IGM attenuations by short
and long horizontal arrows. We note that the IGM attenuations for other
bands in Figure~6 are zero (for $R$ and $i'$) or negligibly small (for
$V$). Table~5 gives a summary of the IGM attenuation.

\begin{table}
 \caption[]{A summary of reddenings by IGM and dust for a source at $z=3.09$.}
 \setlength{\tabcolsep}{3pt}
 \footnotesize
 \begin{minipage}{\linewidth}
  \begin{tabular}{llcc}
   \hline
   & Colour excess & $E({\rm NB359}-R)$ & $E(V-i')$ \\
%   \hline
%   IGM         & Smallest 0.15\%  & 0.35 & --- \\
%   (II08)      & Median           & 1.08 & --- \\
   \hline
   IGM         & Smallest 0.15\%  & 0.28 & --- \\
   (updated)    & Median           & 0.60 & --- \\
   \hline
   Dust$^a$   & Calzetti         & 0.90 & 0.20 \\
               & SMC              & 0.20 & 0.84 \\
   \hline
  \end{tabular}

  $^a$These colour excesses correspond to $E(B-V)=0.1$ in the rest-frame.
 \end{minipage}
\end{table}%

\subsection{Dust attenuation}

If galaxies contain dust, they are reddened. Let us assume two types of
dust attenuation law: the Calzetti law \citep{cal00} and the
extinction law of the Small Magellanic Cloud (SMC). The former
attenuation law was derived from spectra of local UV-selected starburst
galaxies \citep{cal00} and is routinely assumed in studies of high-$z$
galaxies. However, recent studies for $z=2$--3 LBGs suggest that a
steeper attenuation law than the Calzetti one is favourable for some
cases \citep{red06,sia09}. Thus, we also adopt the SMC law which is the
steepest among the known attenuation/extinction laws. We use the model
by \cite{wei01} which reproduces the empirical SMC extinction law very
well.

The Calzetti law is originally only for $\lambda\ge1200$ \AA\
\citep{cal00}. However, we simply extrapolate the original formula for
$\lambda<1200$ \AA. This could be justified by the study of \cite{lei02}
up to $\lambda=970$ \AA\ (but see also \citealt{bua02}) and also be 
justified by the model of \cite{wei01} which does not show any breaks or
features up to about 700 \AA\ for graphite and astronomical silicate
\citep{dra84}. Table~5 gives a summary of the dust reddenings.

In Figure~6, we show the region which can be explained by a model with
normal sub-solar metallicity and IMF and with a combination of dust
and IGM attenuations. While a half of LBGs are found in the region, all
the LAEs are out of it. We need a new model to explain these LAEs.

\subsection{Models with escaping nebular Lyman continuum}

Paper I has proposed the importance of an additional contribution by
escaping nebular LyC. For an escape of the stellar LyC, the ionized
nebulae should be `matter-bounded', at least along some lines of sight, 
from which we can also expect an escape of nebular LyC produced by the
recombination process. Paper I shows that this escaping nebular LyC
makes a peaky spectral feature just below the Lyman limit, which we call
Lyman limit `bump' (see Figs.~3 and 4 in Paper I). In fact, our NB359
filter exactly captures this Lyman limit bump at $z\simeq3.1$. Let us
compare this scenario with the observed strength of LyC.

Figure~9 shows how the Lyman limit bump blues NB359$-R$ colour; 
the colour becomes bluer when a fraction of LyC is absorbed by
nebulae and reemitted in the Lyman limit bump than when all the LyC
escapes. Indeed, the bluest colour is realized when about 40\% LyC is
absorbed (i.e. the escape fraction is about 60\%), irrespective of the
stellar population model. Based on Figure~9, we find that moderate
sub-solar metallicity and normal IMF (solid line: model A1) cannot reach
NB359$-R<0.4$ (0.7) for the minimum (median) IGM attenuation even if we
consider the additional contribution by nebular LyC. On the other hand,
some $z\simeq3.1$ LAEs detected in LyC show a much bluer colour.
Furthermore, even the LAEs with NB359$-R\geq0.4$ are difficult to
be explained because of their red colour in $V-i'$ which expect much
redder NB359$-R$ as shown in Figure~6.

\begin{figure}
 \begin{center}
  \includegraphics[width=7cm]{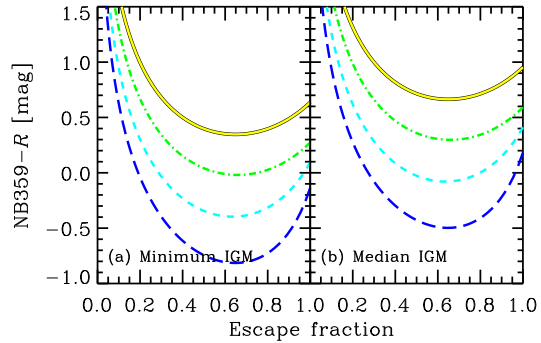}
 \end{center}
 \caption{NB359$-R$ colour for $z=3.1$ galaxies, which corresponds to
 the strength of the Lyman limit `bump', as a function of the escape
 fraction of stellar LyC: (a) minimum IGM attenuation case and (b)
 median IGM attenuation case. The curves are the models with escaping
 nebular LyC: the long-dashed curve for the model D in Table~4, the
 short-dashed curve for the model C, the dot-dashed curve for the model
 B1, and the solid curve for the model A1. The nebular gas temperature
 is assumed to be $1\times10^4$ K.}
\end{figure}

\begin{figure}
 \begin{center}
  \includegraphics[width=7cm]{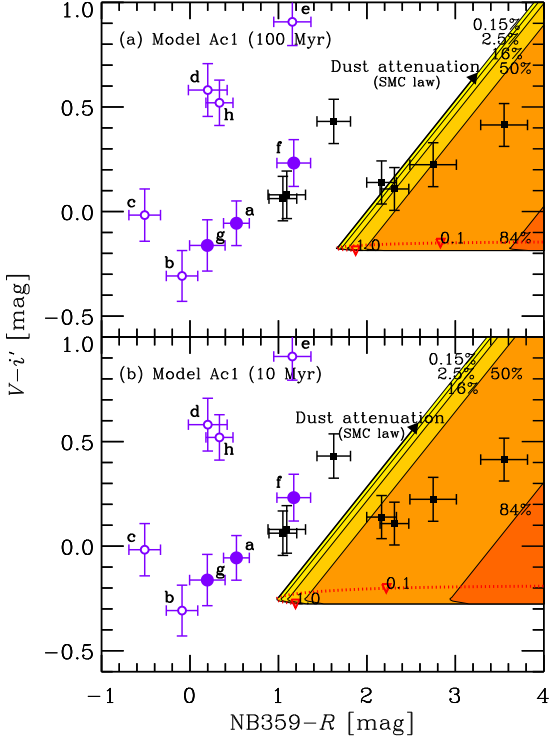}
 \end{center}
 \caption{Same as Fig.~6 but a comparison with the escaping nebular LyC
 scenario of a continuous star-forming galaxy with $Z=1/50\,Z_\odot$
 (model Ac1): (a) age from the onset of star formation of 100 Myr and
 (b) 10 Myr. The dotted curves indicate colour sequences as a function
 of the escape fraction of stellar LyC with the minimum IGM attenuation.
 The symbols indicate the positions with the escape fraction of 1.0,
 0.1, or 0.01. The shaded regions indicate the regions explained with a
 combination of IGM and dust attenuations. From thin to thick, the
 cumulative probability to have the amount of the IGM attenuation
 increases. The SMC extinction law is adopted for the dust attenuation
 as shown by the solid arrow. The nebular gas temperature is assumed to
 be $1\times10^4$ K.}
\end{figure}

\begin{figure}
 \begin{center}
  \includegraphics[width=7cm]{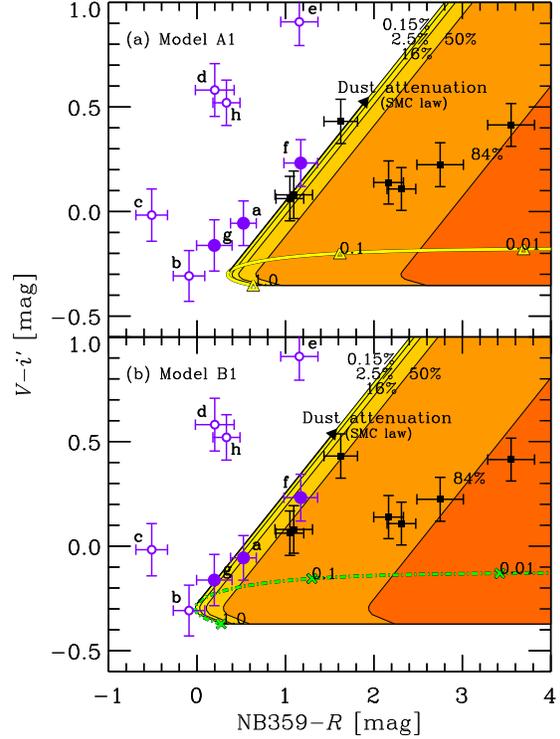}
 \end{center}
 \caption{Same as Fig.~10 but the cases of a very young (1 Myr) stellar
 population with $Z=1/50\,Z_\odot$: (a) normal IMF (model A1; solid line
 with triangles) and (b) extremely top-heavy IMF (model B1: mean stellar
 mass of $\sim50$ $M_\odot$; dot-dashed line with crosses).}
\end{figure}

\begin{figure}
 \begin{center}
  \includegraphics[width=7cm]{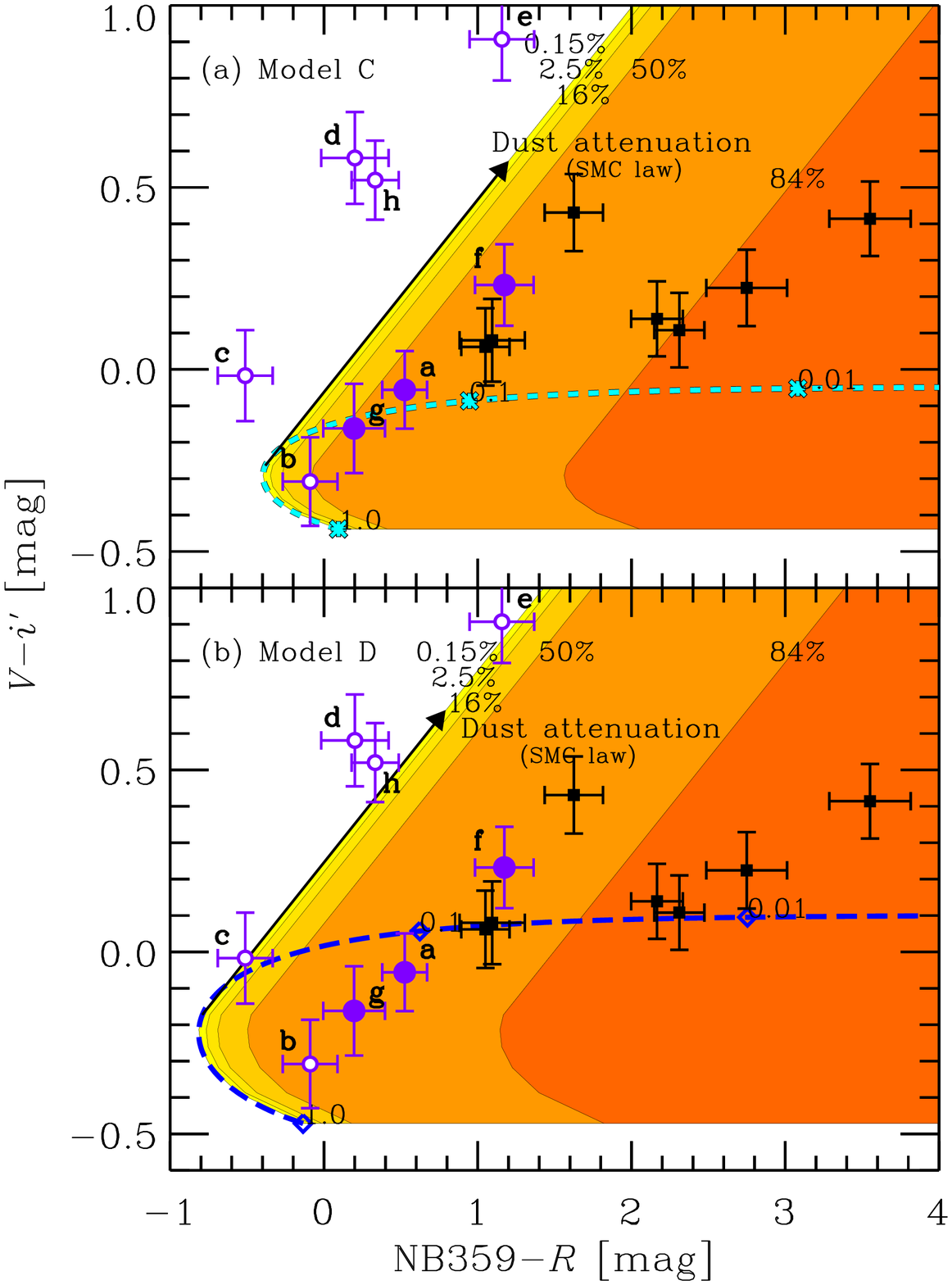}
 \end{center}
 \caption{Same as Fig.~10 but the cases of a very young (1 Myr) and
 massive ($\sim100$ $M_\odot$) stellar population with (a)
 $Z=1/2000\,Z_\odot$ (model C; short-dashed line with asterisks) and (b)
 $Z=0$ (model D; long-dashed line with diamonds).} 
\end{figure}

Figures~10--12 show the two-colour diagrams same as Figure~6 but with
escaping nebular LyC models. In each panel, the curve with symbols shows
the colour sequence as a function of the escape fraction of the stellar
LyC. We have assumed the minimum IGM attenuation for the curves. The
symbols indicate the colours when the escape fraction is 1.0 (i.e. pure
stellar colour + minimum IGM), 0.1, or 0.01. As the escape fraction
decreases, NB359$-R$ colour first becomes bluer than the stellar one due
to the Lyman limit bump as shown in Figure~9, and then, the colour turns
over when the escape fraction is about 0.6 and becomes redder and redder
after the bluest point. In the same time, $V-i'$ colour becomes
redder than the stellar one because of the nebular continuum by
two-photon and bound-free processes. The amount of these colour changes
depends on the strength of the stellar LyC: larger changes by stronger
LyC. 

The shaded region in each panel of Figures~10--12 shows the colours
explained by a combination of IGM and dust attenuations for a specific
stellar population model. The thickness of the shades indicate the
cumulative probability of the IGM attenuation: higher probability,
thicker. We have assumed the SMC extinction law for the dust attenuation
but the Calzetti law cases are included in the shaded regions
like in Figure~6.

Figure~10 shows the cases of the normal stellar population with a
constant star formation (Ac1 in Table~4). We have assumed
$Z=1/50\,Z_\odot$, very low, but, sometimes observed metallicity
(hereafter we call it normally sub-solar). 
The case with $Z=1/5\,Z_\odot$ is always redder in NB359$-R$ than
the case with $Z=1/50\,Z_\odot$. We find that all the LAEs
(filled and open circles) and half of the LBGs (squares) cannot be
explained even with the duration of the star formation of 10--100 Myr
which is younger than a typical age of LBGs ($\sim300$ Myr;
\citealt{sha03}). Therefore, we conclude that the LyC emitting LAEs
should have a stellar population much younger than 10 Myr or much more
massive than the standard Salpeter IMF.

Figure~11 shows the cases of very young (age of 1 Myr) stellar
populations with $Z=1/50\,Z_\odot$, normally sub-solar metallicity. 
The case with $Z=1/5\,Z_\odot$ is about 0.2 mag redder in NB359$-R$
than the case. 
Assuming the standard Salpeter IMF (panel [a]; model A1), we find that
it is still difficult to explain all the LAEs. On the other hand, if we
assume an extremely top-heavy IMF whose mean mass is about 50 $M_\odot$
(panel [b]; model B1), the LAEs without line offset (filled circles) can
be explained with an attenuation smaller than the median. However, it is
still difficult to explain the LAEs with line offset (open circles). We
may have a few true LyC emitters in the 5 objects because we have
statistically rejected the possibility that all the 5 have a faint
foreground object accounting for the NB359 flux in \S4.2.2. There is
another difficulty of the model B1. The critical metallicity for the IMF
change from the standard to top-heavy, $Z_{\rm cr}\la10^{-3}Z_\odot$
\citep{bro03,schn03,sch06} is expected to be much lower than
$Z=1/50\,Z_\odot$ which is assumed in the model B1. 
Thus, a very massive IMF under the
relatively `high' metallicity in the model B1 is unlikely. Then, we
conclude that it is difficult to explain the strength of LyC of the LAEs
with normally sub-solar metallicity unless the critical metallicity for
the IMF change is much higher than expected in the literature.

Figure~12 shows the cases of very young (age of 1 Myr) and massive
($\sim100$ $M_\odot$) stellar populations with $Z=1/2000\,Z_\odot$ (EMP;
model C; panel [a]) and $Z=0$ (Pop III; model D; panel [b]). In these
cases, we can expect a very massive IMF according to the critical
metallicity scenario \citep{bro03,schn03,sch06}. With the model C, we can
easily explain the three LAEs which are confirmed to be true LyC
emitters, whereas the LAEs with line offset are still difficult, except
for the object {\bf b}. The Pop III case (model D) can explain the
objects {\bf b} and {\bf c}, and marginally explain {\bf e} and {\bf h}.
The object {\bf d} is still difficult but this object shows the largest
offset of the Ly$\alpha$ (Table~2), and thus, the foreground probability
is the highest. On the other hand, the object {\bf c} shows the smallest
offset and is the most likely object as a real LyC emitter among the 5
offset LAEs. Enough interestingly, the Pop III model can explain the
object with a 1--20\% smallest IGM attenuation. In summary, very young
($\sim1$ Myr) and massive ($\sim100$ $M_\odot$) EMP stars can reproduce
the observed colours of the three LAEs without line offset and Pop III
stars can reproduce even the colours of most of the LAEs with line
offset. {\it Because a few line-offset LAEs are likely to be real LyC
emitters, we conclude that the Pop III model is the most favourable
stellar population for the LyC emitting LAEs.}

\subsection{Models with two stellar populations}

\begin{figure}
 \begin{center}
  \includegraphics[width=7cm]{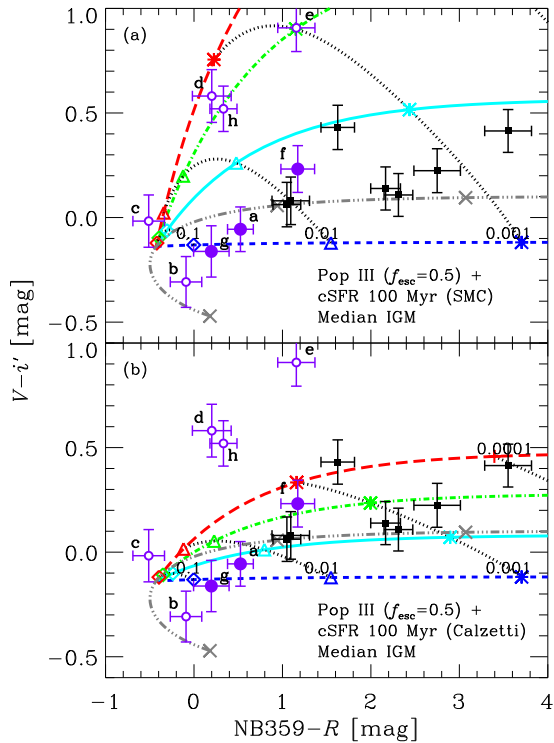}
 \end{center}
 \caption{Same as Fig.~6 but comparisons with models of two stellar
 populations: the cases with models D (Pop III) and Ac2 (underlying
 population with a sub-solar metallicity, $1/5\,Z_\odot$, and a normal
 IMF) in Table~4. A constant star formation of 100 Myr is assumed for
 the underlying population. The SMC extinction law is assumed for the
 panel (a) but the Calzetti attenuation law is assumed for (b). The
 median IGM attenuation is assumed for both panels. The short-dashed
 curves are the sequences of the colour as a function of the mass
 fraction of Pop III stars for dust-free underlying stellar
 population. The solid, dot-dashed, and long-dashed curves are the same
 sequences but for dusty underlying population with $E(B-V)=0.1$, 0.2,
 and 0.3, respectively. Note that the Pop III stars are assumed to be
 always dust-free. The positions for the mass fraction of 0.1, 0.01,
 0.001, and 0.0001 are indicated by dotted curves with squares,
 triangles, asterisks, and plus-mark, respectively. For Pop III stars,
 the contribution of escaping nebular LyC is taken into account,
 assuming the escape fraction of 0.5. The colour sequence with other
 escape fractions is shown by the triple-dot-dashed curve as in
 Fig.~12. For the underlying population, the escape fraction of 0.01 is
 assumed.}
\end{figure}

The previous subsection shows that the LAEs detected in LyC by I09
are likely to contain a primordial stellar population such as massive
Pop III or EMP stars. But how much amount of these exotic stars are
required in them? Also, are Pop III stars really compatible with a
slight dust attenuation which is required to explain some LAEs with line
offset and red in $V-i'$ in Figure~12? To discuss these questions, let
us consider a system in which a primordial stellar population and a
normal stellar population with sub-solar metallicity and Salpeter IMF
coexist. We assume that the normal population makes stars with a
constant rate; the model Ac2 in Table~4 is adopted for this population. 
For the primordial population, we adopt the model D (Pop III). The
stellar mass of the normal population is defined as the multiple of the
star formation rate and the duration; we neglect the loss of the stellar
death for simplicity. The mass fraction of the primordial population in
the total stellar mass is a parameter for the mixture. The contribution
of nebular continuum in LyC and in other wavelengths is taken into
account for both populations but with different escape fractions: 0.5
for the primordial population and 0.01 for the normal
population.\footnote{In addition to the photo-ionized gas, we can
consider the free-free radiation from hot gas produced by SNe and
stellar winds. We have assumed that the total energy of the hot gas
radiation to be 5\% and 1\% of the bolometric stellar luminosities for
the primordial population (as done in Paper I) and for the normal
population \citep{lei99}, respectively. The hot gas temperature is
assumed to be $1\times10^6$ K. However, this hot gas contribution is
negligible as shown in Fig.~3 of Paper I.} The normal population may
exist in dusty environment, but the primordial population are likely to
be dust-free. We adopt two dust attenuation laws: the SMC extinction law
and the Calzetti attenuation law as described in \S5.3 only for the
normal underlying population.

The required mass fraction of the primordial stellar population depends
on the duration of the star formation in the underlying normal
population. This is because the mass of the primordial population which
is required to explain blue NB359$-R$ colour does not change very
much because the colour is determined mainly by the primordial
population, while the mass of the underlying population is proportional
to the duration. For example, we consider 100 Myr as a duration for
$z=3$ galaxies \citep[e.g.,][]{sha03}. The results are shown in
Figure~13. For other durations, the primordial mass fraction is roughly
estimated by an inverse relation to the durations of the underlying star
formation. Although we assumed the median IGM attenuation in Figure~13,
the readers can shift all the model curves along the horizontal axis for
other IGM attenuations.

The LAEs without offset (objects {\bf a}, {\bf f} and {\bf g}; filled
circles in Fig.~13) require a primordial mass fraction of 0.1--10\%
depending on the IGM and dust attenuations as well as the star formation
duration. The dust attenuation for the underlying population is modest
as $E(B-V)=0$--0.1 for the SMC case or up to 0.3 for the Calzetti case,
which is reasonable for $z\sim3$ LAEs \citep[e.g.,][]{gaw06}. The bluest
LAEs (objects {\bf b} and {\bf c}; open circles) require a primordial
mass fraction of more than 10\%, but we could not reject the possibility
that their NB359 flux was foreground contamination individually
(\S4.2). The LAEs which are red in $V-i'$ (objects {\bf d}, {\bf e} and 
{\bf h}; open circles) can be explained by a model with the SMC law and
a small amount of attenuation as $E(B-V)\simeq0.2$--0.3 and with a
primordial mass fraction of a few 0.1\%. When the Calzetti law is
adopted, the red LAEs require a large amount of dust attenuation as 
$E(B-V)>0.5$ which may be too large. Note that the possibility which the
NB359 flux of these objects, especially {\bf d} and {\bf e}, are
foreground contamination is the largest because of the largest line
offset (\S4.2).

The LBGs from which I09 reported detections of LyC (filled squares) are
found in the range of the primordial mass fraction of 0.01--1 \%
depending on the dust attenuations. Thus, only a very small amount of
primordial stars can account for the observed LyC of these galaxies. On
the other hand, some of these LBGs can be explained by a normal
population with a large escape fraction of $\sim0.5$ as shown in
Figures~10 and 11. Therefore, the primordial stars are not mandatory for
them.

\section{Summary and discussions}

We have investigated the nature of the LAEs at $z\simeq3.1$ detected in
our deep narrowband NB359 imaging with Subaru/S-Cam. The NB359 captures
LyC from $z>3$. Thus, we detect LyC from the LAEs if they are at
$z\simeq3.1$. These LAEs are special because of their surprisingly strong
LyC relative to non-ionizing UV (I09). Deep follow-up spectroscopy with
VLT/VIMOS and Subaru/FOCAS for 8 such LAEs presented in \S3 shows 
that at least three of them are highly likely to be at $z\simeq3.1$ and
the NB359 captures truly their LyC (\S4.1.1 and \S4.1.2). From the
spectra of the three LAEs, we have derived the rest-frame EW of
Ly$\alpha$ which are not very large as expected from primordial stellar
population models such as Pop III stars. We have also derived upper
limits on the He {\sc ii} $\lambda1640$ emission line which are not
strict enough to reject Pop III star formation (Table~3). Other five out
of the 8 LAEs show a $\sim0.''8$ offset between the Ly$\alpha$ emission
and the continuum detected in NB359 (Figure~1). For these LAEs, we could
not determine the redshifts of the continuum sources (\S4.2.1). However,
it is statistically difficult that all the five have foreground
contamination which explain the NB359 detections (\S4.2.2). Thus, NB359
may capture LyC from a few of the five LAEs with line offset.

Very interestingly, all the LAEs reported by I09 are too bright in LyC
to be explained by normal stellar population models (Figure~6). Although
we reduced IGM opacity to match the latest LLS statistics (Figures 7 and
8), compared to the previous estimate, this is true even for the three
LAEs which are confirmed to be at $z\simeq3.1$. 
Unlike traditional spectral models of galaxies, Paper I proposed a
new model taking an escape of nebular LyC into account. This model
expects a strong flux excess just below the Lyman limit: we call it
Lyman limit `bump'. We tried to explain the observed strength in LyC of
the LAEs with this new Lyman `bump' model. As a result, we have found
that the colour of the LAEs cannot be explained by the stellar
population with the standard Salpeter IMF and normally sub-solar
metallicity ($Z\geq1/50\,Z_\odot$) even with the Lyman `bump'
(Figures~10 and 11a). If we assume a very young ($\sim1$ 
Myr) stellar population with an extremely top-heavy IMF (mean mass of
$\sim50$ $M_\odot$) and $Z=1/50\,Z_\odot$, the three LAEs without line
offset can be explained with a relatively smaller IGM attenuation but
other five LAEs with line offset cannot be explained yet (Figure 11b). 
Because (1) a few of the line-offset LAEs are possibly to be real LyC
emitters and (2) the expected critical metallicity for a top-heavy IMF
is much lower than the assumed $Z=1/50\,Z_\odot$, we reject this model. 
This suggests that the LAEs contain a significant amount of `primordial'
stars such as EMP or Pop III stars.

A very young ($\sim1$ Myr) and massive ($\sim100$ $M_\odot$) EMP model
can reproduce the observed colours of one line-offset LAE as well as the
LAEs without offset (Figure 12a). A Pop III model ($\sim1$ Myr and
$\sim100$ $M_\odot$) can reproduce all but one LAEs discussed here
(Figure 12b). The last LAE (object {\bf d}) shows the largest offset
between NB359 position and Ly$\alpha$, thus, the possibility of a
foreground contamination in NB359 is the highest. Finally, we conclude
that the Pop III model is the most favourable stellar population for the
LyC emitting LAEs although the EMP model may be still compatible. If we
consider a combination of two stellar populations, `primordial' and
`normal' sub-solar metallicity with dust, the UV two-colour diagram of
all the LAEs and LBGs can be well explained (Figure~13). In this case,
the mass fraction of the `primordial' stars is estimated to be 0.1--10\%
or more for LAEs with $V-i'<0.3$, a few 0.1\% for LAEs with $V-i'>0.3$
(but these may be foreground contamination), and 0.01--1 \% for LBGs,
depending on the age of the galaxies, dust content, and IGM attenuation
(Figure~13). We also note that the LBGs can be explained by normal
stellar populations and EMP or Pop III stars are not mandatory for them.

The Lyman `bump' galaxies presented in this paper is a new population of
high-$z$ galaxies. Indeed, they were missed in standard `drop-out'
surveys because they are not `drop-out' but have an excess flux in LyC:
Lyman `bump'. Because we started from a sample selected based on a
narrowband excess by the Ly$\alpha$ emission line, we did find another
excess in LyC. However, \cite{ste01} and \cite{sha06} started from a
LBG sample, i.e. `drop-out' sample, and thus, they could not find any
galaxies with a Lyman `bump'. This new population of galaxies probably
often emits a strong Ly$\alpha$ emission because they should
intrinsically emit strong LyC although a half of the LyC may directly
escape. Therefore, starting from a LAE sample selected by a narrowband
would be the best to sample the Lyman `bump' galaxies. On the other
hand, if all the LAEs had Lyman `bump', i.e. no `drop-out', LAEs and
LBGs would separate well. However, a significant fraction of LBGs
overlap with LAEs in fact \citep[e.g.,][]{sha03,nol04,ver08}. This may
imply that the fraction of LAEs having Lyman `bump' is small although
the IGM attenuation may play a role to hide the Lyman `bump' and to make
`drop-out'.

A caveat is the lack of the confirmation of the Lyman limit `bump' in
spectroscopy. Our argument of finding the Lyman limit `bump' is only
based on narrowband photometry. We should confirm the feature in
spectroscopy in future.

\subsection{A possible evolutionary scenario of Lyman `bump' galaxies}

\begin{figure}
 \begin{center}
  \includegraphics[width=10cm,angle=-90]{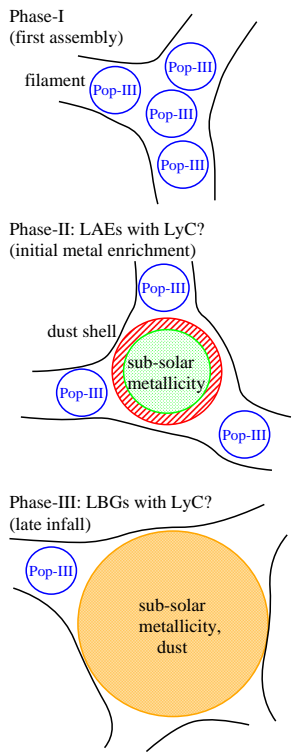}
 \end{center}
 \caption{Schematic picture of an evolutionary scenario of Lyman `bump'
 galaxies.}
\end{figure}

The model with two stellar populations shown in Figure~13
simultaneously reproduces the strength of the Lyman limit `bump' and
non-ionizing UV slope of the LAEs and LBGs detected in LyC by
I09. In the model, we have assumed that a primordial stellar
population which is dust-free coexists with `normal' stars with
sub-solar metallicity and dust. We here discuss a possible scenario
which may justify this assumption. Figure~14 shows a schematic picture
of the scenario.

The hierarchical structure formation scenario argues that a galaxy is
assembled from many subgalactic minihaloes. Cosmological hydrodynamics
simulations have shown that the galaxy assembly occurs through cosmic
filaments and galaxies reside at nodes of the filaments 
\citep[e.g.,][]{bro09}. \cite{gre08} examined the first galaxy assembly
at $z\sim10$ and found that in a $5\times10^7$ $M_\odot$ halo, $\sim10$
very massive metal-free stars, so-called Population III.1, were formed
before the assembly. As a radiative feedback by the Pop III.1 stars,
pristine gas which was ionized once by the Pop III.1 stars can be more
abundant in H$_2$ and HD than before, cool more efficiently, and form
intermediate-mass metal-free stars, so-called Population III.2 
\citep[e.g.,][]{yos07}. As a result, \cite{gre08} expected that gas
infalling onto the central part of a galaxy in the potential well likely
forms Pop III.2 during the assembly if the gas remains primordial. The
Pop III.1 and III.2 stars emit intense LyC which produces cosmological H
{\sc ii} and He {\sc iii} regions. The escape fraction of LyC is
expected to be very high (60--90\%; \citealt{kit04,yos07}) and the
nebular LyC can make a significant Lyman limit `bump' (Phase-I).

While very massive Pop III.1 may become blackhole directly and make no
contribution to the metal enrichment, Pop III.2 stars will end their
life with core-collapse supernovae (SNe) and enrich surrounding gas with
metal \citep[e.g.,][]{heg03}. At the same time, dust grains are also
formed by the SNe \citep[e.g.,][]{noz03}. The metal and dust enrichment
starts from the central part of the galaxy where the mode of star
formation changes into the `normal' mode \citep[e.g.,][]{sch06}. If the
multiple SNe blast waves make a supershell which contains dust grains,
the dust attenuation becomes a screen geometry in which the dust
attenuation law is determined only by dust optical properties. If the
dust properties resemble the SMC which is also making an intense star
formation, the reddening law has a steep wavelength dependence like the
SMC extinction law which is assumed in the upper panel of Figure~13. On
the other hand, primordial gas and Pop III stars may be still infalling
onto the galaxy from the IGM. These infalling primordial stars can make
a significant Lyman limit `bump' (Phase-II).

As the star formation proceeds in the main part of the galaxy, dust and
metal are mixed well in the ISM. In such a medium, the wavelength
dependence of the dust attenuation law becomes weak and expected to be
similar to the Calzetti law \citep{fis03,ino05} which is assumed in the
lower panel of Figure~13. Even in this late stage, a small portion of
the surrounding IGM may remain primordial if the metal enrichment in the
IGM is not efficient. The primordial gas may form Pop III stars during
its infalling process through a cosmic filament (Phase-III).

With this scenario, we can explain the two-colour diagram as follows; 
the LAEs correspond to the Phase-II (or possibly Phase-I). If the age of
the galaxies is 10--100 Myr, the primordial mass fraction is
$\sim1$--10\% or more in some cases. The amount of the dust attenuation
for the underlying population is $E(B-V)=0$--0.3. The LBGs correspond to
the Phase-III because of their relatively large stellar mass 
($\sim10^{10-11}$ $M_\odot$). If the age is 100--1,000 Myr, the
primordial mass fraction is $<1$\% and typically $\sim0.1$\%. The amount
of the dust attenuation is $E(B-V)=0$--0.3 which is similar to LAEs but
the attenuation laws may be different from each other.

We can also explain another observational fact: spatial displacements of
LyC and non-ionizing UV of the LBGs found in I09. As shown in Fig.~2 of
I09, LyC detected position coincides with a substructure found in the
periphery of the main component of the LBGs, although some of these LyC
emitting spots might be foreground contaminations \citep{van10}. In our
scenario, we expect that primordial stars emitting LyC exist in the
periphery of the galaxies (see also \citealt{tor07}). This is exactly
the finding from the LBGs by I09. We also expect that LAEs show a
smaller offset between LyC and non-ionizing UV than LBGs. According to
I09, the offsets of the LAEs are not significant, and thus, smaller than
those found in the LBGs. Again, this is consistent with our scenario.

On the other hand, we have found another spatial displacement between
continuum including LyC and Ly$\alpha$ emission line in some LAEs. 
While these continua may not come from the LAEs but from foreground
objects, the probability that all the LAEs with line offset have
a foreground contamination is very low as seen in \S4.2.2. We have also
proposed a possible mechanism to produce such a line offset in \S4.2.3: 
the LAEs are composed of several unresolved substructures which have
different Ly$\alpha$ EWs. This mechanism may fit in our scenario as
follows: in the Phase-II, the continuum including LyC comes from
substructures in the periphery and Ly$\alpha$ emission comes from the
dusty central component, because the Ly$\alpha$ transmission can be
higher than the continuum transmission if the dust is confined in clumps
in the supershell \citep{neu91}.

\subsection{Primordial stars at $z\sim3$ and intergalactic metal
  enrichment}

Massive ($\sim100$ $M_\odot$) EMP or Pop III stars exist in $z\sim3$
LAEs if our interpretation is correct. Since these primordial stars are
short-lived (a few Myr) because of their mass, pristine gas should
remain to form them until $z\sim3$.  This argument is not consistent
with a scenario of global metal enrichment by Pop III stars at very
early epoch \citep{mac03}, in which the entire universe was enriched to
a metallicity floor as $Z>10^{-3}Z_\odot$ by Pop III stars at $z>20$. 
On the other hand, recent studies suggest inhomogeneous metal enrichment
in the IGM and an extended epoch of Pop III star formation as described
below.

Observations of metal absorbers in high-$z$ QSO spectra show
that metal enrichment in Ly$\alpha$ forest is inhomogeneous
\citep{sim04,bec09}. For example, Simcoe et al.~(2004) show that
$\sim30$\% of lines of sight at $z\sim2.5$ have metallicity lower than
the detection limit of $Z\sim10^{-3}Z_\odot$. Cosmological hydrodynamics
simulations also show that metal enrichment in the IGM proceeds in a
very inhomogeneous way and that a significant volume in the IGM remains
pristine until a lower redshift \citep{tor07,opp09,tre09,wie09,mai10}. 
For example, Oppenheimer et al.~(2009) show that only $\sim1$\% volume
in the IGM is enriched to $Z>10^{-3}Z_\odot$ by $z=5$. Therefore, IGM
metal enrichment is likely to proceed in a patchy way and we can expect
a significant chance to form EMP or metal-free stars from pristine
materials kept until $z\sim3$ \citep{tor07,joh10}. In this sense, our
argument of the discovery of EMP or metal-free stars at $z\sim3$ is
consistent with recent studies of IGM metal enrichment.

The LAEs which may contain primordial stars reside in a massive
proto-cluster region, SSA22 field \citep{ste98,hay04}. This field is
very peculiar in an enhanced activity of LAEs/LABs (M04), dusty sub-mm
galaxies \citep{tam09}, and AGNs \citep{leh09}. Such activities may
proceed IGM metal enrichment efficiently by metal-rich outflows which
are observed ubiquitously around LBGs \citep{ste10}. On the other hand,
the LAEs' clustering in this field is very low \citep{hay04} and this
indicates that the LAEs do not lie in the high density peak of the
proto-cluster but in the less dense periphery of the large-scale
filaments \citep{shi07}. The metal enrichment may not proceed well
yet there. However, the exact three-dimensional distribution of the
LAEs and other types of galaxies by spectroscopy is not obtained yet and
is an interesting future work.

\subsection{Implications for ionizing background and reionization}

The existence of massive EMP or Pop III stars in LAEs may have an
impact on the studies of ionizing background and reionization. 
Based on the observed cosmic evolution of ionizing background,
\cite{ino06} have suggested higher LyC emissivity (or escape
fraction) of galaxies at higher-$z$. Such an evolving LyC emissivity is
supported by galactic LyC observations \citep{sha06,sia10,bri10} and
simulations \citep{raz06,raz10} and is favourable for the cosmic
reionization \citep{bol07}. On the other hand, the physical mechanism of
the evolving LyC emissivity is still uncertain. Some radiative
transfer simulations suggest that less massive galaxies have larger
escape fraction of LyC \citep{raz10,yaj10}. \cite{van10c} have found a
possible luminosity dependence of the LyC emissivity from a large LBG
sample. If such less massive galaxies are common at higher-$z$, the
global LyC emissivity becomes larger at higher-$z$. Another possibility
is that galaxies at higher-$z$ have a higher mass fraction of massive
EMP or Pop III stars and their LyC emissivity is intrinsically
higher. In \S5.5, we have found that more than an order of magnitude
larger primordial mass fraction in the LAEs (1--10\%) than the LBGs
(0.1\%). If this larger primordial mass fraction is true for general
LAEs and the LAE population is more common at higher-$z$ as found by
\cite{ouc08}, the galactic LyC emissivity becomes higher at
higher-$z$. Of course, these two scenarios can be combined: the less
massive LAE population has larger intrinsic LyC emissivity due to
primordial stars and larger escape fraction. Indeed, to make a strong
Lyman `bump', we need a large escape fraction of $\sim0.5$ (Paper
I). Therefore, such less massive Lyman bump galaxies may play a
significant role in the cosmic reionization at $z>6$.

\section*{Acknowledgments}

We are grateful to Jean-Michel Deharveng for suggestions, discussions,
and encouragements, to Hisanori Furusawa for cooperation in the Subaru
observations (S07B-010), and to Eros Vanzella, Masami Ouchi, Masao Mori,
Hidenobu Yajima, Masakazu Kobayashi, Hiroshi Shibai, and Toru Tsuribe
for discussions and comments. We thank anonymous referee for her/his
comments which help us improve the content of this paper.
A.K.I. is supported by KAKENHI (the Grant-in-Aid for Young Scientists B:
19740108) by The Ministry of Education, Culture, Sports, Science and
Technology (MEXT) of Japan and by the Institute for Industrial Research,
Osaka Sangyo University. In the early phase of this work,
A.K.I. was invited to the Laroratoire d'Astrophysique de Marseille and
financially supported by the Observatoire Astronomique de
Marseille-Provence.

\label{lastpage}

\end{document}